\documentclass[11pt]{article}
\usepackage[top=1in,bottom=1in,left=1.1in,right=1.1in]{geometry}
\usepackage[onehalfspacing]{setspace}

\usepackage[english]{babel}
\usepackage[T1]{fontenc}
\usepackage{amsmath}
\usepackage{amssymb}
\usepackage[lighttt]{lmodern}
\usepackage{mathabx}
\usepackage{mathrsfs}
\usepackage{bbm}

\let\int\undefined
\DeclareSymbolFont{otherlargesymbols}{OMX}{cmex}{m}{n}
\DeclareMathSymbol{\int}{\mathop}{otherlargesymbols}{"52}
\let\oldint\int
\renewcommand{\int}{\oldint\nolimits}

\let\sum\undefined
\DeclareSymbolFont{otherlargesymbols}{OMX}{cmex}{m}{n}
\DeclareMathSymbol{\sum}{\mathop}{otherlargesymbols}{"50}

\usepackage{titlesec}
\titleformat{\section}
{\bfseries}
{\thesection}{1em}{\MakeUppercase} 
\usepackage{indentfirst}
\usepackage{titletoc}
\usepackage{tocloft}

\usepackage[bottom]{footmisc}
\usepackage{threeparttable}
\usepackage{booktabs}
\usepackage{graphicx}
\usepackage{subcaption}   
\usepackage{caption}      
\usepackage{cases}
\usepackage[title]{appendix}
\usepackage{enumerate}

\usepackage{amsthm}
\theoremstyle{plain} 
\newtheorem{assumption}{Assumption}
\newtheorem{assumptionS}{Assumption}

\newtheorem{assumptionSPrime}{Assumption}

\newtheorem{assumptionR}{Assumption}
\newtheorem{lemma}{Lemma}
\newtheorem{lemmaA}{Lemma}

\newtheorem{theorem}{Theorem}
\newtheorem{corollary}{Corollary}

\newtheorem{proposition}{Proposition}

\theoremstyle{definition} 
\newtheorem{definition}{Definition}

\newtheorem{example}{Example}
\newtheorem{remark}{Remark}[section]
\newtheorem{myalgorithm}{Algorithm}

\theoremstyle{remark}

\usepackage{natbib}
\usepackage[dvipsnames]{xcolor}
\definecolor{myblue}{RGB}{0,25,150}
\usepackage[colorlinks=true,citecolor=myblue,linkcolor=myblue]{hyperref}

\newcommand{\I}{\mathbbm{1}}
\newcommand{\G}{\mathbb{G}}
\newcommand{\rkd}{\mathrm{RKD}}
\newcommand{\qrkd}{\mathrm{QRKD}}
\newcommand{\drkd}{\mathrm{DRKD}}
\newcommand{\mrkd}{\mathrm{MRKD}}
\newcommand{\pto}[1]{\overset{#1}{\to}}
\newcommand{\wto}[2]{\overunderset{#1}{#2}{\leadsto}}

\begin{document}
\allowdisplaybreaks

\begin{titlepage}
	\onehalfspacing
	\newcommand{\maintitle}{%
	A Unified Framework for Identification and Inference of Local Treatment Effects in Sharp Regression Kink Designs}
	\newcommand{\wzx}{\textsc{Zhixin Wang}}
	\newcommand{\wzxemail}{wangzhixin.ok@qq.com}
	\newcommand{\zzy}{\textsc{Zhengyu Zhang}}
	\newcommand{\zzyemail}{zy.zhang@mail.shufe.edu.cn}
	\newcommand{\sufe}{\normalsize{School of Economics, Shanghai University of Finance and Economics}}
	
	\title{
		\maintitle%
		\footnote{The two authors contributed equally and are listed in alphabetical order. 
			Address correspondence to Zhengyu Zhang, e-mail: \href{mailto:\zzyemail}{\texttt{\zzyemail}}.
			}
		}
	
	\author{\wzx \\ \sufe \vspace{1em}
			\and
			\zzy \\ \sufe}

	\date{ }

	\maketitle

	\begin{abstract}
		This paper develops a unified framework for the identification, estimation, and uniform inference of local treatment effects (LTEs) in sharp regression kink designs (RKDs). These LTEs quantify the effect of a marginal change in the treatment at the kink point on various features of the outcome distribution. The identification strategy applies to Hadamard-differentiable functionals of the outcome distribution---including means, quantiles, and inequality measures---and encompasses several existing RKD estimands as special cases. For estimation, we categorize the corresponding estimands into two general classes and implement their estimation via local polynomial constrained regression. We establish the asymptotic theory for this framework and provide a valid resampling procedure for uniform inference. The method is applied to examine the effect of unemployment insurance on unemployment durations, focusing on the policy's impact on the distribution and inequality of durations, as a complement to existing empirical evidence.

		\vspace{0.5em}
	\end{abstract}
\thispagestyle{empty}
\end{titlepage}

\section{Introduction} 
The regression kink design (RKD) exploits the change in the slope of the relationship between an endogenous treatment variable and a covariate. This quasi-experimental method has become a popular approach in empirical economics, particularly for studying the labor supply effects of social insurance (\citealp{landais2015assessing,card2015effect,kolsrud2018optimal,landais2021value}). Formal econometric frameworks for the RKD are provided by \cite{card2015inference} and  \cite{chiang2019causal}. These frameworks establish that conventional RKD estimands, such as those for means and quantiles, identify a weighted average of structural derivatives, with weights that vary across (sub)populations. However, the causal interpretation of these estimands within the potential outcomes framework (\citealp{rubin2005causal,imbens2015causal}) remains unclear. Furthermore, the identification and inference of more general causal effects in this setting remain relatively underexplored.

This paper develops a unified framework for the identification and estimation of local treatment effects in sharp regression kink designs. Unlike the prior literature, which primarily seeks causal interpretations for specific RKD estimands, our approach begins by defining a broad class of causal parameters of interest and then derives the corresponding strategies for their identification and estimation.
In a sharp RKD, the treatment \(B\) is a deterministic function of the running variable \(X\), given by \(B=b(X)\). The general causal effect at the treatment level \(b_0:=b(x_0)\) is captured by the following parameter:
\begin{align*}
	\Delta_\phi := \frac{\partial}{\partial b} \phi\big(F_{Y(b)|X=x_0}\big)\big|_{b=b_0} 
	:=  \lim_{\delta\to 0}\frac{\phi(F_{Y(b_0+\delta)|X=x_0})-\phi(F_{Y(b_0)|X=x_0})}{\delta}, 
\end{align*}
where \(\phi\) is a functional on the space of one-dimensional distribution functions, and \(Y(b)\) denotes the potential outcome under treatment level \(b\in\mathbb{R}\). Typically, $x_0$ represents the \textit{kink} point in this design, defined as a threshold where the derivative of $b(x)$ is discontinuous: $\lim_{x\to x_0^+}\frac{d}{dx}b(x) \ne \lim_{x\to x_0^-}\frac{d}{dx}b(x)$.
The parameter \(\Delta_{\phi}\) measures the effect of an infinitesimal change in the treatment variable on a feature of the conditional potential outcome distribution $F_{Y(b)|X=x_0}$, evaluated at $b=b_0$. For instance, when \(\phi\) is the mean functional, i.e., \(\phi(F) = \mu(F):=\int w\,dF(w)\), \(\Delta_{\phi}\) corresponds to the ``local average response'' parameter identified in \cite{card2015inference}.

We refer to the parameter \(\Delta_\phi\) as the \textit{local treatment effect} (LTE). This terminology is analogous to that of  \cite{florens2008identification}, who study the identification of the average treatment effect (ATE) in models with a continuous endogenous treatment. In their framework, the ATE at treatment level \(b \in \mathbb{R}\) is defined as
\begin{align*}
	\Delta^\mathrm{ATE}(b) :=\frac{\partial}{\partial b}\mu \big(F_{Y(b)}\big) := \lim_{\delta \to 0} \frac{E[Y(b+\delta)] - E[Y(b)]}{\delta}. 
\end{align*}
In standard settings, the unconditional distribution of a potential outcome, \(F_{Y(b)}\), can often be identified for any \(b\in\mathbb{R}\) using methods like instrumental variables or control functions (\citealp{florens2008identification,imbens2009identification}). This distribution then serves as a foundation for identifying causal parameters such as \(\Delta^\mathrm{ATE}(b)\). However, such approaches are not suitable for the sharp RKD, where the treatment is a deterministic function of the running variable, precluding the existence of a valid external instrument. Instead, identification in a sharp RKD relies on the change in the slope of the treatment assignment function \(b(x)\) at the kink $x_0$. 
While the deterministic nature of treatment assignment is a known restriction, a challenge in identifying the derivative effect $\Delta_\phi$ arises because the conditional distribution of the potential outcome $F_{Y(b_0+\delta)|X=x_0}$ remains unidentified under the kink assumption alone. Consequently, functionals of this conditional distribution, such as its mean or quantiles, cannot be directly recovered.

This paper proposes an identification strategy for the general local treatment effect \(\Delta_{\phi}\) in the sharp RKD, applicable to any Hadamard differentiable functional \(\phi\). The strategy utilizes standard assumptions from the RKD literature, primarily smoothness conditions on relevant structural functions and on the conditional distributions of the unobserved disturbance $\varepsilon$ and the outcome $Y$ given the running variable $X$ (\citealp{card2015inference,chiang2019causal,qu2019uniform}). As a result, we derive a unified identification formula for $\Delta_{\phi}$ and formally establish the relationship between this general class of LTEs and their corresponding conventional RKD estimands.

This paper develops a unified estimation and inference framework for two new classes of RKD estimands, which correspond to the LTEs for various smooth functionals. Motivated by the local polynomial constrained quantile smoother of \cite{chiang2019causal}, we employ a local polynomial constrained regression RKD estimator and establish its asymptotic properties. This approach has two notable features. First, it explicitly imposes the continuity of the conditional distribution $F_{Y|X=x}$ at the kink point $x_0$ as a constraint. Second, unlike methods that require separate local polynomial regressions on each side of the threshold (\citealp{calonico2014robust}), this estimator computes the required left- and right-derivatives in a single step, which can improve computational efficiency. Moreover, we develop uniform inference procedures for the proposed distributional RKD estimators using a multiplier bootstrap method. This inference framework applies to various LTEs. By integrating pivotal methods developed for quantiles by \cite{chiang2019causal}, our framework covers effects on the mean, the distribution, quantiles, and inequality measures such as the Lorenz curve. To complete the methodology, we also provide a bandwidth selector based on mean squared error (MSE) minimization.

To illustrate the applicability of our framework, we analyze the effect of unemployment insurance (UI) benefits on the distribution of unemployment durations using the Continuous Wage and Benefit History (CWBH) data. This analysis extends previous empirical studies, which primarily focused on mean or quantile effects. In particular, we estimate the local distributional treatment effect (LDTE) to assess the marginal impact on the probability distribution, and the local Lorenz treatment effect (LLTE) to examine the effects on the dispersion and inequality of unemployment durations.

The remainder of this paper is organized as follows. The next subsection outlines our contributions to the literature. Section \ref{sec:model.parameters} introduces the model and our parameters of interest. Section \ref{sec:id} establishes our main identification results. Section \ref{sec:id.further} provides further discussion and a detailed comparison with related literature. Section \ref{sec:esti.infer}  develops the estimation and inference framework and derives its asymptotic properties. Section \ref{sec:numerical.illustration}  presents an empirical application and evaluates the finite-sample performance of the estimators via Monte Carlo simulations. Section \ref{sec:conclusion} concludes.
Appendix \ref{sec:apdx} contains the technical assumptions and auxiliary lemmas, while Appendix \ref{sec:apdx.bandwidth} details the bandwidth selection procedure. All mathematical proofs are provided in the online Supplementary Material.

\subsection*{Contribution to the Literature}
This paper contributes to the theoretical literature on regression kink designs (\citealp{card2015inference,dong2018jump,chiang2019causal,chen2019identification,chen2020quantile}). 
The main contribution of this paper is the development of a unified framework that clarifies the causal interpretation of a broad class of sharp RKD estimands within the potential outcomes model. This framework serves two key purposes. First, it provides a unified identification formula for the general causal parameter $\Delta_\phi$. Second, by nesting existing estimands as special cases, it refines and elucidates their causal interpretations. To illustrate the latter, consider the conventional quantile regression kink design (QRKD) estimand:
\begin{align*}
	\qrkd(\tau)
	&:= \frac{\lim_{x\to x_0^+}\frac{\partial}{\partial x} Q_{Y|X}(\tau|x) - \lim_{x\to x_0^-}\frac{\partial}{\partial x} Q_{Y|X}(\tau|x)}{b'(x_0^+) - b'(x_0^-)}
	\\
	&\;= \frac{\lim_{x\to x_0^+} \frac{\partial}{\partial x}  Q_\tau(F_{{Y(b(x))|X=x}}) - \lim_{x\to x_0^-} \frac{\partial}{\partial x}  Q_\tau(F_{{Y(b(x))|X=x}})}{\lim_{x\to x_0^+}\frac{d}{dx}b(x) - \lim_{x\to x_0^-}\frac{d}{dx}b(x)}.
\end{align*}
It is known that $\qrkd(\tau)$ lacks a direct interpretation as a standard quantile treatment effect (QTE). \cite{chiang2019causal} show that $\qrkd(\tau)$ identifies a weighted average of marginal effects across a subpopulation on the boundary $\partial \mathcal{V}(y_\tau,x_0):=\{e\in\mathbb{R}^{d_\varepsilon}:g(b_0,x_0,e)=y_\tau\}$, where $y_\tau:=Q_{Y|X}(\tau|x_0)$.
This paper shows that, under a continuity condition on the conditional density of the outcome $Y$ given the running variable $X$ (Assumption \ref{A:id.smooth.fyx}), the QRKD estimand identifies the local quantile treatment effect (LQTE) for the subpopulation at the kink point $X=x_0$:
\begin{align*}
	\qrkd(\tau)
	= \Delta_Q(\tau)
	:=\lim_{\delta \to 0}\frac{Q_{Y(b_0+\delta)|X}(\tau|x_0) - Q_{Y(b_0)|X}(\tau|x_0)}{\delta}.
\end{align*}  
This continuity assumption, while commonly used for developing asymptotic theory in the RKD and related literature (e.g., \citealp{chiang2019causal,qu2019uniform}), has not previously been applied for identification purposes. The identifying conditions are discussed in greater detail in Section \ref{sec:id.further}.

Second, our estimation and inference framework, based on local polynomial constrained regression, extends the framework of \cite{chiang2019robust} for local Wald estimands. Their approach cannot directly accommodate certain causal parameters, particularly those defined by smooth inequality functionals such as the Lorenz curve and the Gini coefficient, because these parameters lack a local Wald structure. In contrast, our framework addresses this limitation. We show that such LTEs can be expressed as smooth transformations of foundational distributional and quantile RKD estimands. Based on this result, they can be estimated by applying the corresponding transformations to the foundational estimators. Furthermore, the uniform Bahadur representations established for the foundational estimators, combined with the functional Delta method, provide the basis for valid uniform inference on the class of resulting LTEs.

Finally, this paper contributes to the literature on identifying continuous causal effects with endogenous treatments (\citealp{florens2008identification,imbens2009identification,kasy2014instrumental,hoderlein2017corrigendum}). Within this literature, identification often requires recovering the full conditional distribution of potential outcomes. However, this step typically relies on structural assumptions, such as separability (\citealp{florens2008identification}), the existence of control functions (\citealp{imbens2009identification}), or monotonicity conditions (\citealp{kasy2014instrumental,hoderlein2017corrigendum}). Our framework provides an alternative identification approach. We establish a structural equivalence between the LTE and the sharp RKD analogue of the local average structural derivative (LASD) proposed by \cite{hoderlein2007identification,hoderlein2009identification}, as formalized in Lemma \ref{lem:struct.equiv}. This approach is analogous to that of \cite{chernozhukov2015nonparametric}, who exploit a similar connection to a panel version of the LASD to identify quantile derivative effects in nonseparable panel models. Taken together, our work and these related studies suggest that even when identification of the full distribution of potential outcomes is not readily available, causal derivative effects can often be identified by establishing a link to an appropriate version of the LASD.


\section{Model and Target Parameters} \label{sec:model.parameters}
We observe a random sample $\{(Y_i,B_i,X_i)\}_{i=1}^n$, where $Y_i$ is the outcome, $B_i$ is the treatment, and $X_i$ is the running variable. We assume the data are drawn independently and identically distributed (i.i.d.) from a population where these variables are continuous. For notational simplicity, we drop the subscript $i$ when discussing the population model. Let $\mathcal{Y} \subset\mathbb{R}$, $\mathcal{B} \subset \mathbb{R}$, and $\mathcal{X} \subset \mathbb{R}$ denote the supports of $Y$, $B$, and $X$, respectively.

\begin{assumption}[Nonseparable Model]\label{A:model}
	Let $g: \mathbb{R} \times \mathcal{X} \times \mathcal{E} \to \mathbb{R}$ be a measurable function. 			
	\begin{itemize}
		\item[(i)]
		The potential outcome $Y(b)$ for any treatment level $b \in \mathbb{R}$ is given by
		\begin{align*}
			Y(b) &= g(b,X,\varepsilon)
		\end{align*}
		where $\varepsilon \in \mathcal{E} \subset \mathbb{R}^{d_\varepsilon}$ is a vector representing unobserved individual heterogeneity.
		
		\item[(ii)]
		Let the Stable Unit Treatment Value Assumption (SUTVA) hold, which implies $Y = Y(B)$. Given the deterministic treatment assignment rule $B = b(X)$, the observed outcome is thus determined by
		\begin{align*}
			Y = g(b(X),X,\varepsilon).
		\end{align*}
	\end{itemize}
\end{assumption}

Let $I_{x_0}$ be a closed interval containing the kink point $x_0$, and define $I_{x_0}^o:=I_{x_0} \backslash \{x_0\}$.
\begin{assumption}[Sharp Kink Characterization in the First Stage] \label{A:sharp.kink}
	The treatment assignment rule $b(\cdot)$ is continuous on $I_{x_0}$ and continuously differentiable on $I_{x_0}^o$. This implies continuity at the kink point, $b(x_0^+) = b(x_0^-) = b(x_0)$, but a discontinuity in the first-order derivative, $b'(x_0^+) \ne b'(x_0^-)$, where $b'(x_0^\pm):=\lim_{x\to x_0^\pm} \frac{d}{dx}b(x)$.
\end{assumption}

Assumptions \ref{A:model} and \ref{A:sharp.kink} define the nonparametric, nonseparable structural model that forms the basis of our analysis. Assumption \ref{A:model}(i) specifies the general nonseparable form of the potential outcome. Assumption \ref{A:model}(ii) imposes a standard condition from the causal inference literature: the observed outcome $Y$ is determined by the treatment $B$, which itself is generated by a deterministic treatment selection rule. This setup is consistent with the canonical RKD framework (\citealp{card2015inference,chiang2019causal}). Assumption \ref{A:sharp.kink} formally imposes the sharp kink on the treatment selection function, $b(\cdot)$. This requires that $b(x)$ is continuous at the kink point $x_0$, while its first derivative, $b'(x)$, exhibits a jump discontinuity at that point.

Within this model, we focus on the following causal parameter.
\begin{definition}[Local Treatment Effect] \label{def:lte}
	Let $\mathcal{F}$ denote the space of all one-dimensional distribution functions.
	For any functional $\phi: \mathcal{F} \to \mathbb{R}$, the \textit{local treatment effect at the kink} (LTE) is defined as
	\begin{align*}
		\Delta_\phi
		:= \frac{\partial}{\partial b}\phi\big(F_{Y(b)|X=x_0}\big)\big|_{b=b_0}
		:=\lim_{\delta\to 0}\frac{\phi(F_{Y(b_0+\delta)|X=x_0})-\phi(F_{Y(b_0)|X=x_0})}{\delta} 
	\end{align*}
	provided the limit exists, where \(b_0 :=b(x_0)\).
\end{definition}
The LTE parameter $\Delta_{\phi}$ has a clear causal interpretation. Under the structural model defined by Assumptions \ref{A:model}--\ref{A:sharp.kink}, its definition is equivalent to 
$$\lim_{\delta \to 0}\frac{\phi(F_{g(b_0+\delta,x_0,\varepsilon)|X=x_0})-\phi(F_{g(b_0,x_0,\varepsilon)|X=x_0})}{\delta}.$$
This derivative captures the marginal effect of the treatment $B$ at its baseline level $b_0$ on the feature of the potential outcome distribution characterized by the functional $\phi$. The fundamental identification problem, however, is that the key component of this definition—the counterfactual distribution $F_{Y(b_0+\delta)|X=x_0}$—is unobserved, since the structural function $g$ is unknown.

We conclude this section with several key examples of the LTE $\Delta_\phi$.
\begin{example}[Average Effect]\label{exp:average}
	 When $\phi$ is the mean functional, $\phi(F) = \mu(F):=\int w\,dF(w)$, $\Delta_\phi$ becomes the local average treatment effect (LATE):
	\begin{align*}
		\Delta_\mu
		:=\frac{\partial}{\partial b}\mu\big(F_{Y(b)|X=x_0}\big)\big|_{b=b_0}
		:=\lim_{\delta\to 0}\frac{E[Y(b_0+\delta)|X=x_0] - E[Y(b_0)|X=x_0]}{\delta}.
	\end{align*}
\end{example}

\begin{example}[Distributional and Quantile Effects]\label{exp:dist.quantile}
	When $\phi$ is the identity mapping at a point $y$, which we denote by $\phi(F) = Id_y(F) := F(y)$, $\Delta_\phi$ becomes the local distributional treatment effect (LDTE). This measures the marginal effect on the cumulative distribution function:
	\begin{align*}
		\Delta_{Id}(y):= \frac{\partial}{\partial b} Id_y\big(F_{Y(b)|X=x_0}\big)\big|_{b=b_0}:= \lim_{\delta \to 0}\frac{F_{Y(b_0+\delta)|X}(y|x_0) - F_{Y(b_0)|X}(y|x_0)}{\delta}.
	\end{align*} 
	When $\phi$ is the $\tau$th quantile functional, $\phi(F)= Q_\tau(F):=\inf\{w:F(w)\geq \tau\}$ for $\tau \in (0,1)$, $\Delta_\phi$ becomes the local quantile treatment effect (LQTE):
	\begin{align*}
		\Delta_{Q}(\tau) := \frac{\partial}{\partial b}  Q_\tau\big(F_{Y(b)|X=x_0}\big)\big|_{b=b_0}:= \lim_{\delta \to 0} \frac{Q_{Y(b_0+\delta)|X}(\tau|x_0) - Q_{Y(b_0)|X}(\tau|x_0)}{\delta},
	\end{align*}
	where $Q_{Y(b)|X}(\tau|x)$ is a shorthand for $Q_\tau(F_{Y(b)|X=x})$.
\end{example}

\begin{example}[Lorenz Effect]\label{exp:inequality}
	Our framework also covers inequality measures. For instance, if $\phi$ corresponds to the Lorenz curve, $L_\tau(F):=\int_0^\tau Q_u(F)\,du/\mu(F)$, then $\Delta_\phi$ is the local Lorenz treatment effect (LLTE):
	\begin{align*}
		\Delta_{L}(\tau) 
		:= \frac{\partial}{\partial b}L_\tau\big(F_{Y(b)|X=x_0}\big)\big|_{b=b_0} 
		:= \lim_{\delta \to 0}\frac{L_\tau\big(F_{Y(b_0+\delta)|X=x_0}\big) - L_\tau\big(F_{Y(b_0)|X=x_0}\big)}{\delta}.
	\end{align*}
\end{example}

\section{Identification} \label{sec:id} 
The definition of the LTE $\Delta_\phi$ as a derivative makes the class of Hadamard differentiable functionals a natural choice for our analysis. This class is particularly relevant as it includes most statistical measures used in policy evaluation, such as means, quantiles, and various inequality indices.
To formalize this concept, let $\ell^\infty(\mathbb{T})$ denote the space of all bounded functions defined on a set $\mathbb{T}$, and let $\mathbb{B}$ be a Banach space. 
A functional $\phi: \mathcal{F}\subset \ell^\infty(\mathbb{R}) \to \mathbb{B}$ is called \textit{Hadamard differentiable} at $F \in \mathcal{F}$ tangentially to a set $\mathbb{D}_0\subseteq \ell^\infty(\mathbb{R})$, if there exists a continuous linear map $\phi'_F:\ell^\infty(\mathbb{R})\to \mathbb{B}$ such that
\begin{align*}
	\lim_{\delta \to 0} \left\| \frac{\phi(F + \delta h_\delta)- \phi(F) }{\delta} - \phi'_F(h) \right\|_{\mathbb{B}} = 0
\end{align*}
holds for all sequences $\{h_\delta\}$ satisfying $h_\delta \to h\in\mathbb{D}_0$ and $F + \delta h_\delta \in \mathcal{F}$ for all  sufficiently small $\delta$. The map $\phi'_F$ is the Hadamard derivative of $\phi$ at $F$ tangentially to $\mathbb{D}_0\subseteq\ell^\infty(\mathbb{R})$.
Building on this definition, we impose the following assumption on the functional $\phi$.
\begin{assumptionS}[Smooth Functional]\label{A:id.smooth.hadamard}
	The functional \(\phi:\mathcal{F}\to \mathbb{R}\) is Hadamard differentiable at $F_{Y|X=x_0}$, with its Hadamard derivative denoted by $\phi'_{F_{Y|X=x_0}}$.
\end{assumptionS}

The smooth structural model is specified by the following assumption.
\begin{assumptionS}[Smooth Structural Functions]\label{A:id.smooth.g}
	The structural function $g(\cdot,\cdot,e)$ is continuously differentiable for all $e\in\mathcal{E}$, with its partial derivatives with respect to the first and second arguments denoted by $g_1(\cdot,\cdot,e)$ and $g_2(\cdot,\cdot,e)$, respectively.
\end{assumptionS}

Assumption \ref{A:id.smooth.g}, which aligns with Assumptions 1(ii) and 2 in \cite{card2015inference}, imposes smoothness on the structural function, requiring that the outcome varies continuously with both the treatment and the running variable. Notably, this assumption is weaker than Assumption 2(i) of \cite{chiang2019causal}, as we do not impose continuous differentiability with respect to the unobserved heterogeneity $\varepsilon$.
Under this smoothness condition, the partial derivative of the function $h(x,e):=g(b(x),x,e)$ with respect to $x$ is given by
\begin{align*}
	\frac{\partial}{\partial x} h(x,e) = b'(x) g_1(b(x),x,e) + g_2(b(x),x,e).
\end{align*}
A key implication follows: although the structural partial derivatives, $x\mapsto g_1(b(x),x,e)$ and $x\mapsto g_2(b(x),x,e)$, are continuous at $x_0$, the function $x\mapsto h(x,e)$ is not continuously differentiable at this point because $b'(x)$ has a jump discontinuity.

Our identification strategy requires the following smoothness conditions on the conditional distributions given the running variable.
\begin{assumptionS}[Smooth Disturbance Distributions]\label{A:id.smooth.fex}
	The conditional distribution of \(\varepsilon\) given \(X\) is absolutely continuous with respect to Lebesgue measure. Its conditional density, \(f_{\varepsilon|X}(e|\cdot)\), is continuously differentiable on the interval \(I_{x_0}\) for all \(e\in \mathcal{E}\).
\end{assumptionS}

\begin{assumptionS}[Smooth Outcome Distribution]\label{A:id.smooth.fyx}
	The conditional distribution of \(Y\) given \(X\) is absolutely continuous with respect to Lebesgue measure. Furthermore,  for each \(\tau \in (0,1) \), the conditional density \(f_{Y|X}(Q_{Y|X}(\tau|\cdot)|\cdot)\) is continuous and strictly positive on $I_{x_0}$.
\end{assumptionS}

Assumption \ref{A:id.smooth.fex} is used to address selection bias in the identification strategy and is analogous to Assumption 2(iii) in \cite{chiang2019causal}.
A set of sufficient conditions for Assumption \ref{A:id.smooth.fex} is provided in Assumption \ref{A:id.smooth.card2015} (Section \ref{sec:id.further}), which follows the formulation used by \cite{card2015inference}.

Assumption \ref{A:id.smooth.fyx} facilitates the derivation of the unified identification formula. A contribution of this paper relates to the use of this assumption for identification purposes. In the existing regression discontinuity design (RDD) and RKD literature, the continuity of the conditional density $f_{Y|X}$ has been used primarily for inference---specifically, to derive the asymptotic properties of estimators (e.g., \citealp{qu2019uniform,chiang2019causal}). In contrast, our framework is the first to exploit this condition to achieve point identification. Finally, it is important to note that these two assumptions are not redundant, as the continuity of $f_{\varepsilon|X}$ (Assumption \ref{A:id.smooth.fex}) does not, in general, imply the continuity of $f_{Y|X}$.

The following lemma forms the cornerstone of our identification strategy, establishing a key link between the LTE and the causal parameter defined by the structural model. All regularity conditions, prefixed with `R', are collected in Appendix \ref{sec:apdx}.
\begin{lemma}[Structural Representation]\label{lem:struct.equiv}
	Suppose Assumptions \ref{A:model}--\ref{A:sharp.kink}, \ref{A:id.smooth.hadamard}--\ref{A:id.smooth.g}, \ref{A:id.smooth.fyx}, and \ref{A:id.regular}(i)--(ii) hold. Then,
	\begin{align*}
		\Delta_{\phi} 
		= \phi'_{F_{Y|X=x_0}}\left(E\left[-f_{Y|X}(\cdot|x_0)\,g_1(b_0,x_0,\varepsilon) \middle| Y=\cdot, X=x_0\right] \right).
	\end{align*}
\end{lemma}
Lemma \ref{lem:struct.equiv} establishes that the LTE $\Delta_\phi$ can be expressed as the Hadamard derivative of $\phi$ applied to a conditional expectation. The term inside this expectation is the structural derivative of interest, $g_1$, weighted by the negative conditional density of the outcome, $-f_{Y|X}$. 
This representation provides the link to identification. While the LTE is defined using unobserved potential outcomes, the lemma expresses it in terms of two key components: (i) an estimable weight function, and (ii) the conditional expectation of the structural derivative, $E[g_1(b_0,x_0,\varepsilon)|Y=\cdot,X=x_0]$. The second component, which is the object of identification, is analogous to the \textit{local average structural derivative} (LASD) proposed by  \cite{hoderlein2007identification,hoderlein2009identification}.

Building on Lemma \ref{lem:struct.equiv}, the following theorem establishes our main identification result for the LTE.
\begin{theorem}[Identification of the LTE] \label{thm:id.lte}
	Under Assumptions \ref{A:model}--\ref{A:sharp.kink}, \ref{A:id.smooth.hadamard}--\ref{A:id.smooth.fyx}, and \ref{A:id.regular}, the local treatment effect is identified as:
	\begin{align*}
		\Delta_{\phi} 
		= \phi'_{F_{Y|X=x_0}}\left(\drkd(\cdot)\right),
	\end{align*}
	where $\drkd(\cdot)$ is the distributional RKD (DRKD) estimand, defined by:
	\begin{align*}
		\drkd(y):= \frac{\frac{\partial}{\partial x} F_{Y|X}(y|x_0^+) - \frac{\partial}{\partial x} F_{Y|X}(y|x_0^-)}{b'(x_0^+) - b'(x_0^-)}. 
	\end{align*} 
	
\end{theorem}
Theorem \ref{thm:id.lte}  shows that the LTE, $\Delta_{\phi}$, is identified by applying the Hadamard derivative of the functional $\phi$ in the direction of the DRKD estimand. This DRKD estimand is a local Wald-type ratio constructed from the derivatives of the observable conditional distribution on either side of the kink.
Furthermore, this result provides a causal interpretation for a specific case of the generalized local Wald ratio estimand proposed by \cite{chiang2019robust}, defined as
\begin{align*}
	\begin{split}
		&\tau^{(v)}\left(\theta''; \varUpsilon, \varphi, \psi\right) \\
		&\;:= \varUpsilon\left( \frac{\varphi\left(\lim_{x\to x_0^+}\frac{\partial^v E[h_1(Y,\cdot)|X=x]}{\partial x^v}\right)(\cdot) - \varphi\left(\lim_{x\to x_0^-}\frac{\partial^v E[h_1(Y,\cdot)|X=x]}{\partial x^v}\right)(\cdot)}{\psi\left(\lim_{x\to x_0^+}\frac{\partial^v E[h_2(B,\cdot)|X=x]}{\partial x^v}\right)(\cdot) - \psi\left(\lim_{x\to x_0^-}\frac{\partial^v E[h_2(B,\cdot)|X=x]}{\partial x^v}\right)(\cdot)} \right)(\theta''). 
	\end{split}  
\end{align*}
To establish the connection, setting the tuning parameters to $v = 1$, $(\varUpsilon, \varphi ,\psi)=(\phi'_{F_{Y|X=x_0}}, Id, Id)$, $h_1(Y,\cdot) = \I(Y\leq \cdot)$, and $h_2(B,\cdot)=B=b(X)$ yields the following form for their generalized estimand:
\begin{align*}
	\tau^{(1)}\left(\theta''; \phi'_{F_{Y|X=x_0}}, Id, Id\right) = \phi'_{F_{Y|X=x_0}}\left( \frac{\frac{\partial}{\partial x} F_{Y|X}(\cdot|x_0^+) - \frac{\partial}{\partial x} F_{Y|X}(\cdot|x_0^-)}{b'(x_0^+) - b'(x_0^-)}\right)(\theta'')
\end{align*}
Therefore, under the conditions of our theorem, the estimand $\tau^{(1)}(\theta''; \phi'_{F_{Y|X=x_0}}, Id, Id)$ admits a causal interpretation as the local treatment effect, $\Delta_{\phi}$.

We conclude this section by applying Theorem \ref{thm:id.lte} to the functionals introduced earlier. For detailed derivations, we refer interested readers to Appendix S.1 in the supplementary material.
\setcounter{example}{0}
\begin{example}[Average Effect, Continued]
	The mean functional, $\mu(F) = \int w\,dF(w)$, is linear. Consequently, its Hadamard derivative is the functional itself: $\mu'_F(h) = \int w\,dh(w)$. Substituting this into the general formula from Theorem \ref{thm:id.lte} and applying integration by parts yields the familiar RKD estimand for the mean effect: 
	\begin{align}
		\Delta_\mu =  \frac{\frac{d}{dx}E[Y|X=x_0^+] - \frac{d}{dx}E[Y|X=x_0^-]}{b'(x_0^+) - b'(x_0^-)} =:\mrkd. \label{eq:id.late}
	\end{align}
	This result is consistent with the causal interpretation for the MRKD estimand in \cite{card2015inference}, which we compare in detail in Section \ref{sec:causal.mrkd}.
\end{example}

\begin{example}[Distributional and Quantile Effects, Continued]
	For the identity mapping \(\phi = Id_y\), Theorem \ref{thm:id.lte} directly yields the identification of the local distributional treatment effect (LDTE):
	\begin{align*}
		\Delta_{Id}(y) = \frac{\frac{\partial}{\partial x} F_{Y|X}(y|x_0^+) - \frac{\partial}{\partial x} F_{Y|X}(y|x_0^-)}{b'(x_0^+) - b'(x_0^-)} =:\drkd(y) 
	\end{align*}
	for all \( y\in \mathcal{Y}_{x_0}\), where $\mathcal{Y}_{x}$ denotes the support of the conditional distribution $F_{Y|X=x}$. This result provides a direct causal interpretation for the DRKD estimand: it identifies the marginal effect of the treatment on the potential outcome's conditional distribution $F_{Y(b)|X=x_0}$ at the baseline treatment level $b_0$.
	
	For the quantile functional $\phi = Q_\tau$, Hadamard differentiability is guaranteed under certain regularity conditions (see, e.g., Lemma 21.4 of \cite{van2000asymptotic}). Applying Theorem \ref{thm:id.lte} then yields the identification of the local quantile treatment effect (LQTE): 
	\begin{align}
		\Delta_Q(\tau) 
		=\frac{\frac{\partial}{\partial x} Q_{Y|X}(\tau|x_0^+) - \frac{\partial}{\partial x} Q_{Y|X}(\tau|x_0^-)}{b'(x_0^+) - b'(x_0^-)}
		=:\qrkd(\tau) \notag 
	\end{align}
	for all \(\tau \in (0,1)\). 
	This result is a key contribution of our paper. It establishes that  $\qrkd(\tau)$ has a direct and intuitive causal interpretation as the marginal effect on the potential outcome quantile. 
	This contrasts with the complex weighted average interpretation in \cite{chiang2019causal}, which we discuss in detail in Section \ref{sec:causal.qrkd}.
\end{example}

\begin{example}[Lorenz Effect, Continued]\label{exp:id.inequality.effects}
	Our framework extends naturally to inequality measures. Consider the Lorenz curve functional, $L_\tau(F):=\int_0^\tau Q_u(F)\,du/\mu(F)$. Under standard regularity conditions (e.g., Proposition 2 in \cite{bhattacharya2007inference}), its Hadamard derivative is given by:
	\begin{align*}
		[L_\tau]'_{F}(h)
		& = \frac{\int_{0}^{\tau}[Q_u]'_F(h)\,du}{\mu(F)} - \frac{L_\tau(F)}{\mu(F)}\cdot\mu(h).		
	\end{align*}
	Substituting this derivative into the general identification result from Theorem \ref{thm:id.lte} yields the local Lorenz treatment effect (LLTE):
	\begin{align}
		\Delta_{L}(\tau) 
		&= \frac{1}{\mu_0} \left(\int_{0}^{\tau} \qrkd(u)\,du - L_{Y|X}(\tau|x_0) \cdot \mrkd \right)  \label{eq:id.lorenz.te}
	\end{align}
	for all $\tau \in (0,1)$, 
	where $\mu_0:=E[Y|X=x_0]$ and $L_{Y|X}(\tau|x_0):=L_{\tau}(F_{Y|X=x_0})$ is the baseline conditional Lorenz curve. This outcome illustrates the structure of the framework: the treatment effect on the Lorenz curve (LLTE) is identified as a function involving the effects on the mean (MRKD) and quantiles (the integrated QRKD).
\end{example}

\section{Comparison and Further Discussion} \label{sec:id.further}
This section presents a detailed comparison between our main identification results and the existing literature on RKD designs. We focus in particular on the foundational contributions of \cite{card2015inference} and \cite{chiang2019causal}, and conclude by discussing several extensions of our identification framework.

\subsection{Average Effect}\label{sec:causal.mrkd}
\cite{card2015inference} propose a set of sufficient conditions for the causal interpretation of the mean RKD estimand. To facilitate a comparison with our framework, we now summarize their key identifying assumptions—specifically, those required in addition to our shared Assumptions \ref{A:model}--\ref{A:sharp.kink} and \ref{A:id.smooth.g}—along with their main identification result.
\setcounter{assumptionS}{3}
\begin{assumptionSPrime}\label{A:id.smooth.card2015}~
	\begin{itemize}
		\item[(i)]
		The support \(\mathcal{E}\subset I_\varepsilon\) is bounded for some large compact set \(I_\varepsilon\subset \mathbb{R}^{d_\varepsilon}\).
		
		\item[(ii)]
		The set \(\mathcal{A}_\varepsilon:=\{e\in\mathbb{R}^{d_\varepsilon}:f_{X|\varepsilon}(x|e) > 0,\, \forall x \in I_{x_0}\}\) has a positive measure under $\varepsilon$: \(\int_{\mathcal{A}_\varepsilon}\,dF_{\varepsilon}(e) > 0 \).
		
		\item[(iii)] 
		The conditional density \(f_{X|\varepsilon}(\cdot|e)\) is continuously differentiable on \(I_{x_0}\) for all \(e \in I_\varepsilon\).
	\end{itemize}
\end{assumptionSPrime}

\begin{lemma}[\citealp{card2015inference}, Proposition 1]\label{lem:causal.mrkd.card15}
	Suppose Assumptions \ref{A:model}--\ref{A:sharp.kink}, \ref{A:id.smooth.g}, and \ref{A:id.smooth.card2015} hold. Then,
	\begin{align*}
		\mrkd 
		= E\left[g_1(b_0,x_0,\varepsilon)\middle|X=x_0\right] 
		= \int \omega_{x_0}(e) \cdot g_1(b_0,x_0,\varepsilon) \,dF_\varepsilon(e), 
	\end{align*}
	where \(\omega_x(e):=f_{X|\varepsilon}(x|e)/f_X(x)\).
\end{lemma}
We first note that our main conclusion is consistent with \cite{card2015inference}. Under standard regularity conditions that permit interchanging limits and expectations, our LTE for the mean simplifies to:
\begin{align*}
	\Delta_\mu = E\left[g_1(b_0,x_0,\varepsilon)\middle|X=x_0\right]  .
\end{align*}
This is precisely the parameter identified by \cite{card2015inference}, confirming that both frameworks target the same causal object for the mean effect.

Despite this consistency, our identifying assumptions differ in two key respects. First, embedding the mean case within our unified framework requires Assumptions \ref{A:id.smooth.hadamard} and \ref{A:id.smooth.fyx}. While not strictly necessary for identifying the mean effect in isolation, these assumptions are needed for the generality of a framework covering all Hadamard differentiable functionals.
Second, our framework utilizes Assumption \ref{A:id.smooth.fex}---which directly requires the continuous differentiability of the conditional density $x \mapsto f_{\varepsilon|X}(e|x)$---to establish the causal interpretation of $\mrkd$ as the LATE, $\Delta_\mu$. 
While Assumption \ref{A:id.smooth.card2015} provides sufficient conditions for \ref{A:id.smooth.fex} (as shown in their Lemma 2), it is specifically employed by \cite{card2015inference} to enable the population-weighted average interpretation by addressing the measure-zero conditioning problem. In particular, \ref{A:id.smooth.card2015} allows one to rigorously handle conditioning on the measure-zero event $\{X=x_0\}$ and to express $\Delta_\mu$ as a weighted average of the structural derivative over the entire distribution of unobserved heterogeneity. The resulting weight function, $\omega_{x_0}(e)$, is thus well defined and reflects the relative likelihood that an individual with type $\varepsilon=e$ is located at the kink point (\citealp{card2015inference}, Remark 1).

\subsection{Quantile Effect}\label{sec:causal.qrkd}
We now provide a detailed comparison between our framework and that of \cite{chiang2019causal}. The two approaches differ significantly in their starting point. Our analysis addresses a ``forward problem'': we first define a causal effect of interest ($\Delta_Q$) and then derive an identification strategy for it. In contrast, \cite{chiang2019causal} address an ``inverse problem'': they start with the conventional estimand $\qrkd(\tau)$, and investigate what causal interpretation, if any, it admits. To facilitate this comparison, we first summarize their main assumptions and results

The identification strategy of \cite{chiang2019causal}  relies on geometric concepts. Let $h(x,e):=g(b(x),x,e)$ be the reduced-form outcome function. They define a volume in the space of unobservables as $\mathcal{V}(y,x):=\{e\in\mathbb{R}^{d_\varepsilon}:h(x,e) \leq y\}$, with its boundary denoted by $\partial \mathcal{V}(y,x):=\{e\in\mathbb{R}^{d_\varepsilon}:h(x,e)=y\}$. Using the Hausdorff probability measure developed in \citet{sasaki2015quantile}, they derive a reduced-form expression for the key object of their analysis: the conditional quantile partial derivative, $\frac{\partial}{\partial x}Q_{Y|X}(\tau|x)$. The Hausdorff probability measure on the boundary $\partial\mathcal{V}(y,x)$ is defined differently depending on the dimension $M$ of the unobserved heterogeneity. For \(M > 1\), the measure is defined as the ratio of weighted surface areas:  
\begin{align*}
	P^{M-1}_{y,x}(S)
	:= 	\frac{\int_S\frac{1}{\|\nabla_eh(x,e)\|}\,f_{\varepsilon|X}(e|x)\,dH^{M-1}(e)}{\int_{\partial\mathcal{V}(y,x)}\frac{1}{\|\nabla_eh(x,e)\|}\,f_{\varepsilon|X}(e|x)\,dH^{M-1}(e)}
\end{align*}
for any set \(S\) in the collection of Borel subsets of \(\partial\mathcal{V}(y,x)\). For \(M = 1 \), the boundary $\partial\mathcal{V}(y,x)$ is typically a discrete set of points. The Hausdorff measure $H^0$ becomes the counting measure, yielding the discrete probability measure:
\begin{align*}
	P^0_{y,x}(\{e\}):= \frac{\frac{1}{\|\nabla_eh(x,e)\|}\,f_{\varepsilon|X}(e|x)}{\sum_{e\in\partial\mathcal{V}(y,x)}\frac{1}{\|\nabla_eh(x,e)\|}\,f_{\varepsilon|X}(e|x)}
\end{align*}
for each point $e\in \partial\mathcal{V}(y,x)$.

\setcounter{assumptionS}{4}
\begin{assumptionSPrime}[\citealp{chiang2019causal}, Assumptions 2--3]\label{A:id.smooth.cs19}~
	\begin{itemize}
		\item[(i)]
		The boundary \(\partial\mathcal{V}(y,x)\) is a smooth manifold that can be parameterized by a mapping \(\varPi_{y,x}: \varSigma \to \partial\mathcal{V}(y,x)\) satisfying Assumption 3 of \cite{chiang2019causal}, where \(\varSigma\) is an \((M-1)\)-dimensional rectangle.
		
		\item[(ii)]
		\(e\mapsto h(x,e)\) is continuously differentiable for all \(x\in I_{x_0}\), and  \(\|\nabla_e h(x,e)\| \ne 0\) on \(\partial \mathcal{V}(y,x)\) for all \((y,x)\in \mathcal{Y}_x \times I_{x_0}\)
		
		\item[(iii)]
		\(\int_{\partial \mathcal{V}(y,x)} f_{\varepsilon|X}(e|x)\,dH^{M-1}(e) > 0\) for all \((y,x)\in \mathcal{Y}_x \times I_{x_0}\).
	\end{itemize} 
\end{assumptionSPrime}
We now relate Assumption \ref{A:id.smooth.cs19} to the conditions in \cite{chiang2019causal}. Assumption \ref{A:id.smooth.cs19}(i) is identical to their Assumption 3, while Assumptions \ref{A:id.smooth.cs19}(ii) and (iii) correspond to parts of their Assumption 2. The remaining conditions in their Assumption 2 are already encompassed by our earlier Assumptions \ref{A:id.smooth.g} and \ref{A:id.smooth.fex}. It is worth noting that \ref{A:id.smooth.cs19}(ii) is stronger than Assumption \ref{A:id.smooth.g} in one respect, as it implies the existence and continuous differentiability of the function $e\mapsto g(b(x),x,e)$---a condition not required under \ref{A:id.smooth.g}.

Under these conditions, \cite{chiang2019causal} exploit the sharp kink to derive a reduced-form expression for the QRKD estimand that is free from selection bias. This result is stated in the following lemma.
\begin{lemma}[\citealp{chiang2019causal}, Theorem 1 and Corollary 1]\label{lem:causal.qrkd.cs19}
	Suppose Assumptions \ref{A:model}--\ref{A:sharp.kink}, \ref{A:id.smooth.g}--\ref{A:id.smooth.fex}, and \ref{A:id.smooth.cs19} hold, and that the probability measure \(P^{M-1}_{y,x}\) on \(\partial \mathcal{V}(y,x)\) is well-defined for \((y,x)\in \mathcal{Y}_x \times I_{x_0}\).\footnote{We omit the regularity assumptions that ensure \(P^{M-1}_{y,x}\) is a well-defined probability measure and that justify the use of the dominated convergence theorem. Interested readers may refer to  \cite{chiang2019causal} and \cite{sasaki2015quantile} for details.} Then,
	\begin{align*}
		\qrkd(\tau) 
		 =  E_{P^{M-1}_{y_\tau,x_0}}\left[g_1(b(x_0),x_0,\varepsilon)\right]
		 = \int_{\partial \mathcal{V}(y_\tau,x_0)} g_1(b(x_0),x_0,e)\,dP^{M-1}_{y_\tau,x_0}(e), 
	\end{align*}
	where \(y_\tau:=Q_{Y|X}(\tau|x_0)\).
	If \(\varepsilon\) is a random scalar and \(\partial \mathcal{V}(y_\tau,x_0)\) a singleton, then
	\begin{align*}
		\qrkd(\tau) = g_1\big(b(x_0),x_0,\varepsilon(y_\tau,x_0)\big), 
	\end{align*}
	where \(\varepsilon(y_\tau,x_0)\) denotes the sole element of \(\partial \mathcal{V}(y_\tau,x_0)\).
\end{lemma}
The causal interpretation provided by \cite{chiang2019causal} differs significantly from ours. Their result shows that the QRKD estimand identifies a complex weighted average of the structural derivative, $E_{P^{M-1}_{y_\tau,x_0}}[g_1(b_0,x_0,\varepsilon)]$, where the weighting depends on the geometry of the structural function. In contrast, our Theorem \ref{thm:id.lte} establishes that the same estimand identifies the local quantile treatment effect, $\Delta_Q(\tau)$, a direct marginal effect of the treatment on the potential outcome's conditional quantile $Q_{Y(b)|X}(\tau|x_0)$ at the baseline level $b_0$.
The divergence in results stems from a difference in the underlying identification assumptions. 
Interpreting $\qrkd(\tau)$ as the local quantile treatment effect $\Delta_Q(\tau)$ requires Assumption \ref{A:id.smooth.fyx}. This assumption is less restrictive than Assumption \ref{A:id.smooth.cs19}, which is used in approaches addressing the measure-zero conditioning problem. Notably, Assumption \ref{A:id.smooth.fyx} does not require certain conditions employed in those contexts, such as smoothness of the function $e \mapsto h(x,e)$ or conditions ensuring the Hausdorff probability measure is well-defined.
The ability to obtain a direct causal interpretation under these less restrictive conditions is an advantage of our framework.

To clarify the relationship between Assumption \ref{A:id.smooth.fyx} and more primitive conditions such as Assumption \ref{A:id.smooth.cs19}, we provide two sets of sufficient conditions under which \ref{A:id.smooth.fyx} holds. These conditions involve smoothness restrictions on the conditional density $f_{\varepsilon|X}$ and the structural function $g$.
\begin{assumptionS}\label{A:id.smooth.cs19.f3}
	 Define
	\[\tilde{f}(y,x):=\int_{\partial\mathcal{V}(y,x)} \frac{1}{\|\nabla_eh(x,e)\|}f_{\varepsilon|X}(e|x)\,dH^{M-1}(e). \]
	Suppose \(\tilde{f}(y_\tau(\cdot),\cdot)\) is continuous on \(I_{x_0}\) for all \(\tau \in (0,1)\).
\end{assumptionS}

\begin{assumptionS}\label{A:id.smooth.cherno2015}~
	\begin{itemize}
		\item[(i)]
		The unobserved heterogeneity $\varepsilon$ can be decomposed into two components, $\varepsilon=(\varepsilon_1,A)$, satisfying the following conditions:
		(a) The scalar component $\varepsilon_1$ is absolutely continuous given $(A,X)$. Its conditional density, $f_{\varepsilon_1|A,X}(e_1|a,x)$ is continuous in both $e_1$ and $x \in I_{x_0}$, and is strictly positive for all $a$.
		(b) The vector component $A$ is absolutely continuous given $X$. Its conditional density $f_{A|X}(a|x)$ is continuous in $x \in I_{x_0}$, and is strictly positive for all $a$.
		
		\item[(ii)]
		The function $h(x,e_1,a):=g(b(x),x,e_1,a)$ is continuously differentiable with respect to its second argument $e_1$ for all $a$ and $x \in I_{x_0}$. Its partial derivative $\partial_{e_1}h$ satisfies $\partial_{e_1}h(x,e_1,a) \geq c > 0$ for some constant $c>0$. Furthermore, $x\mapsto \partial_{e_1}h(x,e_1,a)$ is continuous for all $e_1$ and $a$.
		 
	\end{itemize}
\end{assumptionS}

\begin{proposition}\label{pron:relation.density}
	Suppose Assumptions \ref{A:model}--\ref{A:sharp.kink}, and \ref{A:id.smooth.g} hold. Then, 
	Assumption \ref{A:id.smooth.fyx} is satisfied if either of the following two conditions holds:  (a) Assumption \ref{A:id.smooth.cs19} and \ref{A:id.smooth.cs19.f3} hold. (b) Assumption \ref{A:id.smooth.cherno2015} holds.
\end{proposition} 
Proposition \ref{pron:relation.density} provides two distinct sets of primitive conditions under which Assumption \ref{A:id.smooth.fyx} holds. The first set, Condition (a), follows the approach of \cite{chiang2019causal}. It requires the structural function to be smooth with respect to the entire vector of unobserved heterogeneity but does not impose monotonicity. The second set, Condition (b), utilizes an alternative set of assumptions similar to those in \cite{chernozhukov2015nonparametric} and \cite{hoderlein2017corrigendum}. It requires the existence of at least one continuously distributed scalar component within the unobserved heterogeneity vector, and that the structural function is smooth and monotonic with respect to this single scalar component, though not necessarily with respect to others.

\subsubsection{An Alternative Weighted Interpretation}
Our structural representation in Lemma \ref{lem:struct.equiv} establishes a connection between the LQTE and a sharp RKD analogue of the local average structural derivative (LASD) from \cite{hoderlein2007identification,hoderlein2009identification}. Applying Lemma \ref{lem:struct.equiv} with $\phi = Q_\tau$ yields:
\begin{align*}
	\Delta_{Q}(\tau) = E\left[g_1(b_0,x_0,\varepsilon) \big| Y=y_\tau, X=x_0\right].
\end{align*}
While this provides a direct interpretation as a structural marginal effect for a specific subpopulation $\{Y=y_\tau, X=x_0\}$, this interpretation involves conditioning on an event of Lebesgue measure zero. Handling such conditioning rigorously presents known theoretical difficulties (\citealp{sasaki2015quantile}).

Two approaches can address the issue of conditioning on a measure-zero event. The first approach, following \cite{chiang2019causal}, involves replacing our Assumption \ref{A:id.smooth.fyx} with Assumption \ref{A:id.smooth.cs19}. This leads to their weighted-average interpretation on the boundary $\partial\mathcal{V}(y_\tau,x_0)$, as given by Lemma \ref{lem:causal.qrkd.cs19}. However, integrating this approach within our unified framework would necessitate imposing the additional Assumption \ref{A:id.smooth.cs19.f3} to ensure that Assumption \ref{A:id.smooth.fyx} still holds. We propose an alternative approach, showing that the measure-zero issue can be addressed by replacing Assumption \ref{A:id.smooth.fyx} with the following condition instead.
\begin{assumptionS}~\label{A:id.smooth.fyxe}
	\begin{itemize}	
		\item[(i)] 
		The set $\mathcal{A}_\varepsilon(\tau):=\{e\in\mathbb{R}^{d_\varepsilon}:f_{Y,X|\varepsilon}(y_\tau(x),x|e) > 0,\, \forall x\in I_{x_0}\}$ has a positive measure under $F_\varepsilon$ for all $\tau \in (0,1)$: $\int_{\mathcal{A}_\varepsilon(\tau)} \,dF_\varepsilon(e) > 0$.
		
		\item[(ii)]
		The conditional density 
		\(f_{Y,X|\varepsilon}(\cdot,\cdot|e)\) is continuous on \(\mathcal{Y}_x \times I_{x_0}\) for all \(e \in I_\varepsilon\).	
	\end{itemize}
\end{assumptionS}
Assumption \ref{A:id.smooth.fyxe} imposes two conditions on the conditional density of the observables $(Y,X)$ given the unobserved heterogeneity $\varepsilon$. Part (i) requires this density to be strictly positive in a neighborhood of the kink point for a non-trivial subpopulation. Part (ii) requires this density to be continuous in the same neighborhood. Notably, Assumption \ref{A:id.smooth.fyxe}, in conjunction with Assumption \ref{A:id.smooth.card2015}, provides sufficient conditions for Assumptions \ref{A:id.smooth.fex} and \ref{A:id.smooth.fyx}, which enable the derivation of the unified identification formula.
The following proposition provides an alternative weighted causal interpretation for the QRKD estimand.
\begin{proposition}\label{pron:qrkd.lasd.interpret}
	Suppose Assumptions \ref{A:model}--\ref{A:sharp.kink}, \ref{A:id.smooth.g}, \ref{A:id.smooth.card2015}, \ref{A:id.smooth.fyxe} and \ref{A:id.regular} hold. Then, 
	\begin{align*}
		\qrkd(\tau) 
		= E\left[g_1(b_0,x_0,\varepsilon) \middle| Y=y_\tau,X=x_0\right] 
		=\int \omega_{y_\tau,x_0}(e) \cdot g_1(b_0,x_0,e)\,dF_\varepsilon(e) 
	\end{align*}
	for all \(\tau \in (0,1)\), where the weight function is \(\omega_{y,x}(e):=f_{Y,X|\varepsilon}(y,x|e)/f_{Y,X}(y,x)\).
\end{proposition}
This proposition provides two results. The first equality shows that the QRKD estimand corresponds to the sharp RKD version of the local average structural derivative. The second equality addresses the issue of conditioning on the measure-zero event $\{Y=y_\tau,X=x_0\}$ by expressing the LASD as a weighted average of the structural derivative, $g_1$, over the entire population of unobserved heterogeneity. The weight function, $\omega_{y_\tau,x_0}(e)$, is well-defined under Assumptions \ref{A:id.smooth.card2015} and \ref{A:id.smooth.fyxe} and represents the relative likelihood of observing an individual with type $\varepsilon=e$ having characteristics $\{Y=y_\tau, X=x_0\}$.

\begin{remark}[The Role of Monotonicity]
	If the structural function $g(b,x,e)$ is monotonic in a scalar error term $e$, then for any given $(y,x)$, the boundary $\partial\mathcal{V}(y,x)$ consists of the unique point $h^{-1}(y,x)$. In this case, the conditional expectation simplifies to the structural derivative evaluated at this point:
	\begin{align*}
		E\left[g_1(b_0,x_0,\varepsilon) \middle| Y=y_\tau,X=x_0\right]  
		&= E\left[g_1(b_0,x_0,\varepsilon)\middle|h(x_0,\varepsilon)=y_\tau, X=x_0\right] \\
		&= g_1(b_0,x_0,h^{-1}(y_\tau,x_0)).
	\end{align*}
	This result is consistent with the special case in \cite{chiang2019causal}, where the unobserved heterogeneity is one-dimensional ($M=1$) and the boundary $\partial\mathcal{V}(y_\tau,x_0)$ is a singleton.
\end{remark}

The following table summarizes the causal interpretations of conventional RKD estimands as established in the prior literature and in this paper.
\begin{table}[h]
	\centering
	\fontsize{10}{15}\selectfont
	\begin{threeparttable}
		\caption{Causal Interpretations of Classic RKD Estimands}\label{tb:struct.rep}
		\begin{tabular}{l l l l}
			\hline
			 Estimands & Causal Interpretations & Conditions  & Related Literature\\
 			\hline
			\(\drkd(y)\) & \(\Delta_{Id}(y)\) & Assumptions \ref{A:model}--\ref{A:sharp.kink}, \ref{A:id.smooth.g}, \ref{A:id.smooth.fex}, \ref{A:id.smooth.fyx} & This paper \\
			\(\mrkd\) & \(\Delta_\mu=E[\omega_{x_0}(\varepsilon)\cdot g_1(b_0,x_0,\varepsilon)] \) & Assumptions \ref{A:model}--\ref{A:sharp.kink}, \ref{A:id.smooth.g}, \ref{A:id.smooth.card2015}  & \cite{card2015inference} \\
			\(\qrkd(\tau)\) & \(\Delta_Q(\tau)\) & Assumptions \ref{A:model}--\ref{A:sharp.kink}, \ref{A:id.smooth.g}, \ref{A:id.smooth.fex}, \ref{A:id.smooth.fyx} & This paper  \\ 
			 & \(E_{P^{M-1}_{y_\tau,x_0}}[g_1(b_0,x_0,\varepsilon)]\) & Assumptions \ref{A:model}--\ref{A:sharp.kink}, \ref{A:id.smooth.g}, \ref{A:id.smooth.fex},  \ref{A:id.smooth.cs19} & \cite{chiang2019causal} \\
			 & \(E[\omega_{y_\tau,x_0}(\varepsilon)\cdot g_1(b_0,x_0,\varepsilon)]\) & Assumptions \ref{A:model}--\ref{A:sharp.kink}, \ref{A:id.smooth.g}, \ref{A:id.smooth.card2015}, \ref{A:id.smooth.fyxe} & This paper  \\
			 \hline
		\end{tabular}
		\begin{tablenotes}\footnotesize
			\item[1.] Role of Assumption \ref{A:id.smooth.hadamard}  (Hadamard Differentiability): This assumption enables the derivation of the general identification formula in Theorem \ref{thm:id.lte}, applicable to any Hadamard differentiable functional. However, it may not be strictly required when analyzing a specific smooth functional if a direct identification argument is available for that particular case.
			\item[2.] Assumption Hierarchy: Under our baseline model for the sharp RKD (Assumptions \ref{A:model}--\ref{A:sharp.kink} and \ref{A:id.smooth.g}), the following relationships hold between the various smoothness conditions discussed:
			(i) Assumption \ref{A:id.smooth.card2015} $\implies$ Assumption \ref{A:id.smooth.fex}. (ii) Assumption \ref{A:id.smooth.cs19} \& \ref{A:id.smooth.cs19.f3} $\implies$ Assumption \ref{A:id.smooth.fyx}. (iii) Assumptions \ref{A:id.smooth.card2015} \& \ref{A:id.smooth.fyxe}  $\implies$ \ref{A:id.smooth.fex} \& \ref{A:id.smooth.fyx}.
		\end{tablenotes}
	\end{threeparttable}
\end{table}

\section{Estimation and Inference} \label{sec:esti.infer}
\subsection{Estimation Strategy}
Our identification results in Section \ref{sec:id} show that local treatment effects can be expressed either as direct Wald-type ratios or as smooth transformations thereof. This section develops a unified estimation and inference framework to handle both cases. To this end, we define two general classes of RKD estimands.

\textit{Type 1. (Local Wald Ratio Estimand)}: 
The first class of estimands takes the form of a local Wald ratio of derivatives:
\begin{align*}
	\rkd_m(\theta) 
	&:=\frac{m^{(1)}(\theta,x_0^+) - m^{(1)}(\theta,x_0^-)}{b'(x_0^+) - b'(x_0^-)}\\
	&:=\frac{\lim_{x\to x_0^+}\frac{\partial}{\partial x}E\left[\varphi(Y,\theta)\middle|X=x\right] - \lim_{x\to x_0^-}\frac{\partial}{\partial x}E\left[\varphi(Y,\theta) \middle|X=x\right]}{\lim_{x\to x_0^+}\frac{d}{dx}(x) - \lim_{x\to x_0^-}\frac{d}{dx}b(x)} , \quad \theta \in \varTheta.
\end{align*}
Here, $m(\theta,x):=E[\varphi(Y,\theta)|X=x]$ is the conditional expectation of a known transformation $\varphi(\cdot,\theta)$ of the outcome, and $m^{(1)}(\theta,x)$ denotes its first-order derivative with respect to $x$. This foundational class subsumes many standard sharp RKD estimands. For instance, setting $\varphi(y,\theta)=\I(y \leq \theta)$ yields the DRKD estimand ($\rkd_F = \drkd$) for identifying the LDTE. Similarly, setting $\varphi(y,\theta)=y$ yields the mean RKD estimand ($\rkd_\mu = \mrkd$) for identifying the LATE.

\textit{Type 2. (Composite Estimand)}:
The second class of estimands consists of composite functionals of the Type 1 estimands and the quantile RKD estimand:
\begin{align*}
	\rkd_{\psi|m}(\theta')
	&:= \psi\big(\rkd_m(\cdot), \qrkd(\cdot) \big) (\theta'), \quad \theta' \in \varTheta'.
\end{align*}
Here, $\psi$ is a functional defined on a product of function spaces. This class is designed to handle LTEs that are identified via transformations of simpler effects. The local Lorenz treatment effect is a key example. By defining $\psi$ as
\begin{align}
	\begin{split}
		\psi_L(g,h) 
		:=  \frac{1}{\mu_0} \left(\int_{0}^{\tau}h(u)\,du - L_{Y|X}(\tau|x_0) \cdot g\right),
	\end{split} \label{eq:def.psi.lorenz}
\end{align}
and choosing $g=\mrkd$ and $h=\qrkd$, the resulting estimand $\rkd_{\psi_L|\mu}$ identifies the LLTE, $\Delta_L$, where \(\theta' = \tau\) and \(\varTheta' = (0,1)\).

Under our assumptions, the function $x \mapsto E[\varphi(Y,\theta)|X=x]$ is continuous at the kink point $x=x_0$. This continuity holds for cases such as the conditional CDF $F_{Y|X}$. This continuity motivates a convenient estimation strategy that jointly estimates the function's level and its left- and right-hand derivatives within a single procedure. We therefore propose a $p$th-order local polynomial constrained regression estimator. For each $\theta$ in a compact set $\widebar{\varTheta}\subset \varTheta$, we solve the following minimization problem:
\begin{align}
		\hat{\alpha}(\theta) 
		=\arg\min_{\alpha \in \mathbb{R}^{2p+1}}\sum_{i=1}^{n}
		\left(\varphi(Y_i,\theta) - \bar{r}_p\left(X_i-x_0 \right)^\top  \alpha \right)^2
		 K\left(\frac{X_i-x_0}{h_{n,\theta}}\right),\quad  \theta \in \widebar{\varTheta}, \label{eq:esti.constrained.reg}
\end{align}
where \(K(\cdot)\) is a kernel function. 
The vector of regressors, $\bar{r}_p(u):=(1, u\delta_u^+, u\delta_u^-, \dots, u^p\delta_u^+, u^p\delta_u^-)^\top\in \mathbb{R}^{2p+1}$, is constructed to impose the continuity constraint, where \(\delta_u^+:= \I(u\geq 0)\), and \(\delta_u^-:= \I(u<0)\). It includes a single intercept but allows for different coefficients on the polynomial terms to the left ($u<0$) and right ($u\geq 0$) of the kink. 
Then, we obtain
\begin{align*}
	\hat{\alpha}(\theta)= \left( \hat{m}(\theta,x_0), \frac{\hat{m}^{(1)}(\theta,x_0^+)}{1!}, \frac{\hat{m}^{(1)}(\theta,x_0^-)}{1!}, \dots , \frac{\hat{m}^{(p)}(\theta,x_0^+)}{p!}, \frac{\hat{m}^{(p)}(\theta,x_0^-)}{p!} \right)^\top,
\end{align*}
The required function and its left- and right-hand derivatives can thus be obtained as $\hat{m}(\theta,x_0)=\iota_1^\top\hat{\alpha}(\theta)$, $\hat{m}^{(1)}(\theta,x_0^+) = \iota_{2}^\top\hat{\alpha}(\theta)$, and $\hat{m}^{(1)}(\theta,x_0^-) = \iota_{3}^\top\hat{\alpha}(\theta)$, where $\iota_j$ is the $j$th standard basis vector (e.g., \(\iota_2=(0,1,0,\ldots,0)^\top\)).

To estimate the quantile derivatives that form the numerator of the QRKD estimand, we follow \cite{chiang2019causal} and use a local $p$th-order polynomial constrained quantile regression. In our notation, the estimator is:
\begin{align}
	\hat{\beta}(\tau) 
	=\arg\min_{\beta \in \mathbb{R}^{2p+1}}
	\sum_{i=1}^{n}\,
	& \rho_\tau\left(Y_i - \bar{r}_p\left(X_i-x_0 \right)^\top  \beta  \right)K\left(\frac{X_i-x_0}{h_{n,\tau}}\right), \quad \tau \in \mathcal{T},
	 \label{eq:esti.constrained.qr.reg}
\end{align}
where $\mathcal{T}\subset (0,1)$ is a closed interval; $\rho_\tau(u) = (\tau - \I(u<0))u$ is the standard check function. The vector of regressors $\bar{r}_p(u)$ is the same as that used for the mean regression, thereby imposing the continuity of the conditional quantile function $x\mapsto Q_{Y|X}(\tau|x)$ at the kink. Let \(Q^{(\nu)}_{Y|X}(\tau|x):=\frac{\partial ^\nu}{\partial x^v}Q_{Y|X}(\tau|x)\) for an integer \(1 \leq \nu \leq p\). Analogous to the mean regression case, the estimated coefficient vector $\hat{\beta}(\tau)$ provides estimates of the conditional quantile function and its scaled derivatives at the kink:
\begin{align*}
	\hat{\beta}(\tau)= \left( \widehat{Q}_{Y|X}(\tau|x_0), \frac{\widehat{Q}^{(1)}_{Y|X}(\tau|x_0^+)}{1!}, \frac{\widehat{Q}^{(1)}_{Y|X}(\tau|x_0^-)}{1!}, \dots , \frac{\widehat{Q}^{(p)}_{Y|X}(\tau|x_0^+)}{p!}, \frac{\widehat{Q}^{(p)}_{Y|X}(\tau|x_0^-)}{p!} \right)^\top,
\end{align*}
The required quantile function and its left- and right-hand derivatives can thus be obtained as $\widehat{Q}_{Y|X}(\tau|x_0)=\iota_1^\top\hat{\beta}(\tau)$, $\widehat{Q}^{(1)}_{Y|X}(\tau|x_0^+) = \iota_2^\top\hat{\beta}(\tau)$, and $\widehat{Q}^{(1)}_{Y|X}(\tau|x_0^-) = \iota_3^\top\hat{\beta}(\tau)$.

\begin{remark}[On the Constrained Regression Approach]
	Our constrained, one-step estimation approach presents several advantages over the common practice of fitting separate local polynomial regressions on each side of the kink (e.g., \citealp{calonico2014robust,card2015inference}). First, it is computationally efficient, particularly when estimating quantile effects over a large grid of quantiles. Second, it directly incorporates the known continuity of the underlying function (e.g., $m(\theta,x)$ or $Q_{Y|X}(\tau|x)$) at the kink point---the same continuity property used in the asymptotic theory. The concept of a constrained estimation approach has been noted in other contexts, for example by \cite{chen2020quantile} for quantile treatment effects in a fuzzy RKD setting.

	It is important, however, to recognize the limitation of this approach. Our constrained estimators are designed specifically for regression kink designs, where the underlying conditional functions are continuous. They are not applicable to standard RDDs, as the primary identifying assumption in RDDs involves a discontinuity in these functions at the threshold (i.e., $m(\theta,x_0^+) \ne m(\theta,x_0^-)$).
\end{remark}

\begin{remark}
	The conditional quantile function is theoretically strictly increasing in $\tau$. This follows from its derivative, $\frac{\partial}{\partial \tau} Q_{Y|X}(\tau|x_0) = 1/f_{Y|X}(Q_{Y|X}(\tau|x_0)|x_0)$, which is positive under Assumption \ref{A:id.smooth.fyx}. While the true quantile function is monotonic, its empirical counterpart, $\widehat{Q}_{Y|X}(\tau|x_0)$, may fail to exhibit monotonicity over the full range of $\tau$ due to sampling variation. To ensure a properly specified estimate, we apply the standard rearrangement procedure of \citet[Equations (2.1)--(2.2)]{chernozhukov2010quantile}. This procedure transforms the potentially non-monotonic estimate into a monotonically increasing function that preserves the quantile properties.
\end{remark}

We now detail the application of the general estimation framework developed above to the specific local treatment effects discussed previously.
\setcounter{example}{0}
\begin{example}[Average Effect, Continued]\label{exp:esti.late}
	To estimate the LATE, $\Delta_\mu$, we apply our Type 1 estimator, the local polynomial constrained regression defined in (\ref{eq:esti.constrained.reg}). The transformation function is the identity, $\varphi(Y,\theta) = Y$, and the parameter $\theta$ is degenerate. The conditional moment function simplifies to $m(x,\theta) = E[Y|X=x]=:\mu_{Y|X}(x)$.
	The estimated coefficient vector, $\hat{\alpha}$, thus provides estimates of the conditional mean and its derivatives at the kink point:
	\[
	\hat{\alpha} =  \left( \hat{\mu}_0, \frac{\hat{\mu}_{Y|X}^{(1)}(x_0^+)}{1!}, \frac{\hat{\mu}_{Y|X}^{(1)}(x_0^-)}{1!}, \dots , \frac{\hat{\mu}^{(p)}_{Y|X}(x_0^+)}{p!}, \frac{\hat{\mu}^{(p)}_{Y|X}(x_0^-)}{p!} \right)^\top .
	\]
	The required left- and right-hand derivatives can be obtained from $\hat\alpha$, yielding the estimator for the LATE:
	\begin{align}
		\widehat{\Delta}_\mu 
		= \frac{\hat{\mu}_{Y|X}^{(1)}(x_0^+) - \hat{\mu}_{Y|X}^{(1)}(x_0^-)}{b'(x_0^+)-b'(x_0^-)}
		= \widehat{\rkd}_{\mu}.\notag
	\end{align}
\end{example}

\begin{example}[Distributional and Quantile Effects, Continued]\label{exp:esti.ldte.lqte}
	We now detail the estimation of the distributional and quantile effects. First, the LQTE, $\Delta_Q(\tau)$, is estimated directly using the local polynomial constrained quantile regression from (\ref{eq:esti.constrained.qr.reg}). As shown previously, the estimated coefficient vector $\hat{\beta}(\tau)$ yields the required left- and right-hand derivatives, which are then used to form the QRKD estimand:
	\begin{align}
		\widehat{\Delta}_{Q}(\tau) 
		= \frac{\widehat{Q}_{Y|X}^{(1)}(\tau|x_0^+) - \widehat{Q}_{Y|X}^{(1)}(\tau|x_0^-)}{b'(x_0^+)-b'(x_0^-)} = \widehat{\qrkd}(\tau). \notag 
	\end{align}
	Second, the LDTE, $\Delta_{Id}(y)$, is estimated using our Type 1 estimator from (\ref{eq:esti.constrained.reg}), with the transformation function set to the indicator $\varphi(Y,y) = \I(Y\leq y)$. Since the LDTE, $\Delta_{Id}(y)$, is identified pointwise for $y \in \mathcal{Y}_{x_0}$ by $\rkd_F(y)$, a common application involves evaluating this effect at specific quantiles of interest, $y_\tau = Q_{Y|X}(\tau|x_0)$. This yields the estimator for the LDTE:
	\begin{align}
		\widehat{\Delta}_{Id}(\hat{y}_\tau) 
		= \frac{\widehat{F}_{Y|X}^{(1)}(\hat{y}_\tau|x_0^+) - \widehat{F}_{Y|X}^{(1)}(\hat{y}_\tau|x_0^-)}{b'(x_0^+)-b'(x_0^-)} = \widehat{\rkd}_{F}(\hat{y}_\tau). \notag 
	\end{align}
	A feature of the constrained estimation approach is that the necessary components for this evaluation are obtained jointly. The local polynomial quantile regression used to estimate the LQTE also yields an estimate of the quantile level itself, $\hat{y}_\tau = \widehat{Q}_{Y|X}(\tau|x_0)$. The LDTE evaluated at the estimated $\tau$-th quantile is then estimated by substituting $\hat{y}_\tau$ into the DRKD estimator.
\end{example}

\begin{example}[Lorenz Effect, Continued]\label{exp:esti.llte}
	The identification result for the LLTE, $\Delta_L(\tau)$, expresses it as a function of components related to the mean and quantile effects. Specifically, constructing an estimator for $\Delta_L(\tau)$ requires estimates of: the conditional mean ($\mu_0$), the mean RKD estimand ($\widehat{\mrkd}$), the conditional Lorenz curve ($L_{Y|X}(\tau|x_0)$), and the quantile RKD estimand ($\widehat{\qrkd}$).
	Each of these components is available from the estimation procedures described previously. The mean-related terms, $\hat{\mu}_0$ and $\widehat{\mrkd}$, are obtained from the local polynomial regression detailed in Example \ref{exp:esti.late}. The quantile-related terms---namely $\widehat{\qrkd}(u)$ needed for the integral and $\widehat{Q}_{Y|X}(u|x_0)$ needed to construct the estimated Lorenz curve $\widehat{L}_{Y|X}(\tau|x_0)$---are obtained from the local polynomial quantile regression detailed in Example \ref{exp:esti.ldte.lqte}. A plug-in estimator for the LLTE is therefore constructed by substituting these previously obtained estimates into the identification formula:
	\begin{align}
		\widehat{\Delta}_{L}(\tau) 
		= \frac{1}{\hat{\mu}_0} \left(\int_{0}^{\tau} \widehat{\qrkd}(u)\,du - \widehat{L}_{Y|X}(\tau|x_0) \cdot \widehat{\rkd}_{\mu} \right) 
		= \widehat{\rkd}_{\psi_L |\mu}(\tau). \label{eq:esti.llte}
	\end{align}
	The integral in this expression can be computed using standard numerical methods.
\end{example}

\subsection{Asymptotic Theory} \label{apdx:asy.theory}
This section establishes the asymptotic properties of our Type 1 and Type 2 estimators, $\widehat{\rkd}_m(\cdot)$ and $\widehat{\rkd}_{\psi|m}(\cdot)$. We begin by defining the necessary notation. Define the following kernel-dependent constant matrices and vectors:  $\widebar{\varGamma}_p:= \int_{\mathbb{R}} \bar{r}_p(u)\bar{r}_p(u)^\top K(u)\,du$, $\bar{\vartheta}_{p,q}^\pm := \int_{\mathbb{R}_\pm}\bar{r}_p(u) u^{q} K(u)\,du$, and $\widebar{\varPsi}_p^\pm:=\int_{\mathbb{R}_\pm} \bar{r}_p(u)\bar{r}_p(u)^\top K^2(u)\,du$. Let $\varepsilon^m(Y,X,\theta) = \varphi(Y,\theta) - m(\theta,X)$ be the error term from the conditional expectation. We define its conditional covariance function as $\sigma_{\varepsilon^m}(\theta_1,\theta_2|x):=E[\varepsilon^m(Y,X,\theta_1)\varepsilon^m(Y,X,\theta_2)|X=x]$.  
\begin{assumption}~\label{A:asy}
	\begin{itemize}
		\item[(i)] (a) The data \(\{(Y_i,X_i)\}_{i=1}^n\) are i.i.d. copies of random vector \((Y,X)\) on the probability space \((\Omega,\mathcal{F},P)\). (b) The density $f_X(\cdot)$ is continuously differentiable on $I_{x_0}$ and satisfies \(0<f_X(x_0)<\infty\).
		
		\item[(ii)] 
		(a) \(m(\theta,x)\) is Lipschitz continuous at \(x=x_0\) and \(m^{(\nu)}(\theta,\cdot)\) is Lipschitz continuous on \( I_{x_0}\backslash\{x_0\}\) for each \(\theta\in\widebar{\varTheta}\) and \(\nu\in\{1,\ldots,p+1\}\).
		(b) The classes of real-valued functions \(\{y\mapsto \varphi(y,\theta):\theta \in \widebar{\varTheta}\} \) and \(\{x\mapsto m(\theta,x):\theta\in\widebar{\varTheta}\}\) are of Vapnik-Cervonenkis (VC) type with a common integrable envelope \(\chi:\mathcal{Y}\times\mathcal{X}\to\mathbb{R}\) satisfying \(\int |\chi(y,x)|^{2+\eta}\,dF_{Y,X}(y,x) <\infty\) for some \(\eta >0\). (c) The conditional covariance \(\sigma_{\varepsilon^m}(\theta_1,\theta_2|\cdot) \in \mathcal{C}^1(I_{x_0}\backslash\{x_0\})\) and \(\sigma_{\varepsilon^m}(\theta_1,\theta_2|x_0^\pm) <\infty\) for each \(\theta_1, \theta_2 \in \widebar{\varTheta}\).
		
		\item[(iii)]
		(a) The kernel function \(K:[-1,1]\to \mathbb{R}_+\) is bounded and continuous, and the class \(\{x\mapsto K((x-x_0)/h): h>0\}\) is of VC type. (b) \(\widebar{\varGamma}\) is positive definite. (c) \(\widebar{\varPsi}^+\), \(\widebar{\varPsi}^-\), and  \(\int_{\mathbb{R}_\pm} |\bar{r}_p(u) u^q K(u)|\,du\) are finite.

		\item[(iv)]
		The bandwidth $h_{n,\theta}$ is specified as $h_{n,\theta} = \varsigma(\theta)h_n$, where $\varsigma:\varTheta \to (0,\infty)$ is a bounded, Lipschitz continuous function. The baseline bandwidth sequence $\{h_n\}$ is required to satisfy the following conditions as $n\to \infty$:
		(a) $h_n \to 0$ and $nh_n^3 \to \infty$;
		(b) $nh_n^{2p+3} \to 0$.
		
	\end{itemize}
\end{assumption}
Assumption \ref{A:asy} collects the regularity conditions required to establish the uniform asymptotic expansion and weak convergence of our estimators. We briefly discuss each component. For Condition (i), Part (a) is the standard i.i.d. sampling assumption; Part (b) imposes smoothness on the density of the running variable, $f_X$. This is a standard condition used to rule out endogenous sorting around the kink point, and can be derived from the more primitive Assumption \ref{A:id.smooth.card2015}. For Condition (ii), Part (a) imposes smoothness on the conditional moment function, $m(\theta,x)$. Continuity at the kink $x_0$ and differentiability away from the kink are the definitional feature of the RKD. Parts (b) and (c) are technical conditions on the complexity (i.e., VC-type) of the relevant function classes, and are analogous to Assumptions (ii)(a) and (ii)(c) in \cite{chiang2019robust}.
Condition (iii) imposes standard requirements on the kernel function, such as boundedness, which ensures a finite asymptotic variance. It is analogous to Assumption 1(iv) of \cite{chiang2019robust} and notably rules out the use of the Gaussian kernel.
Condition (iv) specifies the admissible rates for the bandwidth sequence $h_n$. Part (a) imposes a rate consistent with the RKD literature (\citealp{card2015inference,chiang2019causal}) and standard theory for local polynomial estimation of first-order derivatives. Part (b) is a standard undersmoothing requirement, which ensures that the bias term of the estimator is asymptotically negligible relative to the variance term.

\begin{lemma}[Asymptotic Properties of \(\hat{m}^{(1)}(\cdot,x_0^\pm)\)]\label{lem:asy.m1.hat}~
	\begin{itemize}
		\item[(i)](Uniform Bahadur Representation) Suppose Assumptions \ref{A:asy}(i)--(iii), and (iv)(a) hold.  Then,
		\begin{align*}
		&\begin{bmatrix}
			\sqrt{nh_{n,\theta}^3}\left(\hat{m}^{(1)}(\theta,x_0^+) - m^{(1)}(\theta,x_0^+) 
			- h_{n,\theta}^p \frac{\iota_2^\top  \widebar{\varGamma}_p^{-1} \left(m^{(p+1)}(\theta,x_0^+) \bar{\vartheta}_{p,p+1}^+ + m^{(p+1)}(\theta,x_0^-)\bar{\vartheta}_{p,p+1}^-\right)}{(p+1)!} \right) 
			\vspace{.25em} \\
			\sqrt{nh_{n,\theta}^3}\left(\hat{m}^{(1)}(\theta,x_0^-) - m^{(1)}(\theta,x_0^-) 
			- h_{n,\theta}^p \frac{\iota_3^\top  \widebar{\varGamma}_p^{-1}\left(m^{(p+1)}(\theta,x_0^+) \bar{\vartheta}_{p,p+1}^+ + m^{(p+1)}(\theta,x_0^-)\bar{\vartheta}_{p,p+1}^-\right) }{(p+1)!} \right)
		\end{bmatrix}\\
		& =
		\begin{bmatrix}
			\mathbb{Z}^m_n(\theta,2) \\
			\mathbb{Z}^m_n(\theta,3)
		\end{bmatrix} 
		+ o_{P}(1) 
	\end{align*}
	as \(n \to \infty\) uniformly in \(\theta \in \widebar{\varTheta} \), where 
	\begin{align*}
		\mathbb{Z}^m_n(\theta, k) 
		:= \sum_{i=1}^{n} \frac{\iota_k \widebar{\varGamma}_p^{-1} \bar{r}_p\left(\frac{X_i-x_0}{h_{n,\theta}}\right) K\left(\frac{X_i-x_0}{h_{n,\theta} }\right)\varepsilon^m(Y_i,X_i,\theta)}{\sqrt{nh_{n,\theta}}f_X(x_0)}, \quad k \in \{2,3\}.
	\end{align*}

	\item[(ii)] (Weak Convergence) Suppose Assumption \ref{A:asy} holds. Then,
	\[
	\begin{bmatrix}
		\mathbb{Z}^m_n(\cdot, 2) \\
		\mathbb{Z}^m_n(\cdot, 3)
	\end{bmatrix}
	\leadsto 
	\begin{bmatrix}
		\mathbb{Z}^m(\cdot, 2)\\
		\mathbb{Z}^m(\cdot, 3)
	\end{bmatrix},
	\]
	where \(\mathbb{Z}^m: \varOmega \to \ell^{\infty}(\widebar{\varTheta} \times \{2,3\}) \) is a tight zero-mean Gaussian process with covariance function 
	\begin{align*}
		E\left[\mathbb{Z}^m(\theta_1, k_1) \mathbb{Z}^m(\theta_2, k_2)\right]
		= \frac{\iota_{k_1}^\top\widebar{\varGamma}_p^{-1}\widebar{\varXi}^m_p(\theta_1,\theta_2) \widebar{\varGamma}_p^{-1}\iota_{k_2}}{ f_X(x_0)}
	\end{align*}
	for all \(\theta_1, \theta_2 \in \widebar{\varTheta}\) and \(k_1, k_2 \in \{2,3\}\). The matrix \(\widebar{\varXi}^m_p\) in the covariance is given by \(\widebar{\varXi}^m_p(\theta_1,\theta_2) :=\widebar{\varPsi}_p^+(\theta_1,\theta_2)\sigma_{\varepsilon^m}(\theta_1,\theta_2|x_0^+) + \widebar{\varPsi}_p^-(\theta_1,\theta_2)\sigma_{\varepsilon^m}(\theta_1,\theta_2|x_0^-)\), where
	\begin{align*}
		\widebar{\varPsi}_p^\pm(\theta_1,\theta_2)
		&:=\frac{1}{\sqrt{\varsigma(\theta_1)\varsigma(\theta_2)}}
		\int_{\mathbb{R}_\pm} \bar{r}_p\left(\frac{u}{\varsigma(\theta_1)}\right) \bar{r}_p\left(\frac{u}{\varsigma(\theta_2)}\right)^\top  K\left(\frac{u}{\varsigma(\theta_1)}\right) K\left(\frac{u}{\varsigma(\theta_2)}\right) \,du.
	\end{align*}
	\end{itemize}
\end{lemma}
Part (i) of the lemma establishes the uniform Bahadur representation for our estimator and provides the explicit expression for the asymptotic bias. 
In contrast to standard local $p$th-order polynomial estimators (e.g., \citealp{chiang2019robust}, Lemma 1), the bias term exhibits two specific characteristics. First, it depends on the $(p+1)$th-order derivatives from both sides of the kink. Second, the associated constant matrices $\widebar{\varGamma}_p$ and vectors $\widebar{\vartheta}^\pm_{p,p+1}$ are constructed using the regressor vector $\bar{r}_p(u) \in \mathbb{R}^{2p+1}$ rather than the standard polynomial basis $r_p(u) \in \mathbb{R}^{p+1}$. Part (ii) of the lemma establishes the weak convergence of the normalized processes. This result requires the use of an undersmoothing bandwidth satisfying $nh_n^{2p+3}=o(1)$. This condition ensures that the higher-order bias term from Part (i) is asymptotically negligible. Consequently, the normalized process converges weakly to a zero-mean Gaussian process, free of asymptotic bias.

We now introduce the assumptions required to establish the asymptotic properties of the quantile-based and composite estimands, $\widehat{\qrkd}$ and $\widehat{\rkd}_{\psi|m}$.
\begin{assumption}[\citealp{chiang2019causal}, Assumption 6]~\label{A:asy.cs19}
	\begin{itemize}
		\item[(i)] 
		Assumption \ref{A:asy}(i).
		
		\item[(ii)]
		(a) \(f_{Y|X}(Q_{Y|X}(\cdot|x_0)|x_0)\) is Lipschitz continuous on \(\mathcal{T}\). (b) There exists finite constants \(f_L>0\), \(f_U>0\), and \(\delta>0\) such that for all \(\tau\in\mathcal{T}\), \(|\eta|\leq\delta\) and \(x\in I_{x_0}\), \(f_{Y|X}(Q_{Y|X}(\tau|x)+\eta|x) \in [f_L,f_U]\).

		\item[(iii)]
		(a) \(Q_{Y|X}(\cdot|x_0)\), \(\partial Q_{Y|X}(\cdot|x_0^\pm)/\partial\tau\) exist and are Lipschitz continuous on \(\mathcal{T}\). (b) \(Q_{Y|X}(\tau|\cdot)\) is continuous at \(x_0\); the function \((x,\tau)\mapsto \partial^v Q_{Y|X}(\tau|x)/\partial x^v\) exists and is Lipschitz continuous on \((x,\tau)\in (I_{x_0}\backslash\{x_0\})\times\mathcal{T}\) for all \(v \in \{0,1,\ldots,p+1\}\).

		\item[(iv)]
		The kernel \(K\) is compactly supported, nonnegative, and satisfies \(K'(u) < \infty\), \(\int K(u)\,du = 1\) and \(\int uK(u)\,du = 0\). The matrix  \(\widebar{\varGamma}_p\) is positive definite.

		\item[(v)]
		The bandwidths \(h_{n,\tau}\) satisfies \(h_{n,\tau} = c(\tau)h_n\), where \(nh_n^3 \to \infty\) and \(nh_n^{2p+3} \to 0\) as \(n\to \infty\) and \(c(\tau)\) is Lipschitz continuous satisfying \(0 < \underline{c} \leq c(\tau) \leq \bar{c} <\infty \) for all \(\tau\in \mathcal{T}\).
		
	\end{itemize}
\end{assumption}

\begin{assumption}~\label{A:asy.psi.esti.hadamard}
	\begin{itemize}
		\item[(i)] 
		For each \((g, h)\in \ell^{\infty}(\widebar{\varTheta}) \times \ell^{\infty}(\mathcal{T})\),
		\(\hat{\psi}(g,h)(\cdot) = \psi(g,h)(\cdot)  + o_{P}(1)\) uniformly on \( \widebar{\varTheta}'\).
		
		\item[(ii)]
		\(\psi\) is Hadamard differentiable at \((\rkd_m, \qrkd)\) with derivative \(\psi'_{(\rkd_m, \qrkd)}(\cdot,\cdot)\).
	\end{itemize}
\end{assumption}
Assumption \ref{A:asy.cs19} is required to establish the asymptotic properties of $\widehat{\qrkd}$. This assumption is consistent with Assumption 6 of \cite{chiang2019causal}, and a result that relies on it is restated for completeness as Lemma \ref{lem:asy.qrkd.cs19} in Appendix \ref{sec:apdx}.

Assumption \ref{A:asy.psi.esti.hadamard} provides the high-level conditions required for the asymptotic analysis of our Type 2 composite estimators, $\widehat{\rkd}_{\psi|m}$. Part (i) requires the uniform consistency of any `nuisance' estimators that appear in the definition of the functional $\psi$. For example, when estimating the LLTE, the functional $\psi_L$ depends on the baseline mean $\mu_0$ and the conditional quantile function $Q_{Y|X}(\cdot|x_0)$. The consistent estimators for these nuisance components are conveniently provided by the same constrained regressions, (\ref{eq:esti.constrained.reg}) and (\ref{eq:esti.constrained.qr.reg}), used for the main effects. 
Part (ii) imposes the standard Hadamard differentiability condition on the functional $\psi$. This condition enables the application of the functional Delta method to derive the asymptotic distribution of the composite estimator (e.g., \citealp{kosorok2008introduction}, Theorem 2.8). 

\begin{theorem}[Weak Convergence of RKD Estimators]~\label{thm:asy}
	\begin{itemize}
		\item[(i)] 
		Suppose Assumptions \ref{A:sharp.kink} and \ref{A:asy} hold. Then,
		\begin{align*}
			&\sqrt{nh_{n,\theta}^3}\left(\widehat{\rkd}_m(\theta) - \rkd_m(\theta)\right)
			\leadsto 
			\G^m(\theta):= \frac{\mathbb{Z}^m(\theta,2) - \mathbb{Z}^m(\theta,3) }{b'(x_0^+)-b'(x_0^-)} 
		\end{align*}
		uniformly in \(\theta \in \widebar{\varTheta}\). The limiting process $\G^m(\cdot)$ is a zero-mean Gaussian process with covariance function
		\begin{align*}
			E\left[\G^m(\theta_1)\G^m(\theta_2)\right] 
			= \frac{(\iota_{2}-\iota_{3})^\top\widebar{\varGamma}_p^{-1}\widebar{\varXi}^m_p(\theta_1,\theta_2) \widebar{\varGamma}_p^{-1}(\iota_{2}-\iota_{3})}{(b'(x_0^+)-b'(x_0^-))^2 f_X(x_0) },
		\end{align*}
		where the terms \(\mathbb{Z}^m(\cdot,\cdot)\) and \(\widebar{\varXi}^m_p(\cdot,\cdot)\) are as defined in Lemma \ref{lem:asy.m1.hat}.		
		
		\item[(ii)]
		Suppose Assumptions \ref{A:sharp.kink}, \ref{A:asy}, \ref{A:asy.cs19}, and \ref{A:asy.psi.esti.hadamard} hold, Then
		\begin{align*}
			\sqrt{nh_{n}^3}\left(\widehat{\rkd}_{\psi|m}(\theta') - \rkd_{\psi|m}(\theta') \right) 
			\leadsto
			\G^{\psi|m}(\theta')
			:=\psi'_{(\rkd_m, \qrkd)}\left(\frac{\G^m(\cdot)}{\sqrt{\varsigma^3(\cdot)}},\frac{\G^Q(\cdot)}{\sqrt{c^3(\cdot)}}\right)(\theta')
		\end{align*}
		uniformly in \(\theta'\in\widebar{\varTheta}'\), where $\G^{\psi|m}(\theta')$ is a zero-mean Gaussian process with covariance function \(E\left[\G^{\psi|m}(\theta'_1)\G^{\psi|m}(\theta'_2)\right]\), the form of which depends on \(\psi\). The limiting process \(\G^Q:\varOmega \to \ell^\infty(\mathcal{T})\) is the Gaussian process for the quantile estimator, as defined in Lemma \ref{lem:asy.qrkd.cs19}. 
	\end{itemize}
\end{theorem}
Theorem \ref{thm:asy}(i) provides the weak convergence result for the Type 1 estimators. This result includes the distributional RKD estimator and the mean RKD estimator as special cases. While asymptotic theory for local Wald ratios exists (e.g., the general, higher-order case in \citealp{chiang2019robust}), the result here applies specifically to the first-order derivative and is derived for the constrained estimation approach used in this paper.
Theorem \ref{thm:asy}(ii) establishes the weak convergence for the Type 2 (composite) estimators. One application is the local Lorenz effect estimator. The limiting process $\G^{\psi|m}$ for these estimators is a composite of the limiting processes $\G^m$ and $\G^Q$ for the Type 1 and quantile estimators. Because this limiting process is not a local Wald ratio, the inference framework of \cite{chiang2019robust}, though general in other contexts, cannot be directly applied to these Type 2 estimators.

\subsection{Multiplier Bootstrap and Pivotal Methods}
This section develops a resampling method to approximate the asymptotic distributions derived in Theorem \ref{thm:asy} and to conduct uniform inference.

For the Type 1 estimator, $\widehat{\rkd}_m(\cdot)$,  we employ a multiplier bootstrap to approximate the distribution of the limiting process $\G^m(\cdot)$. The bootstrap process is constructed based on the leading term of the uniform Bahadur representation established in Lemma \ref{lem:asy.m1.hat}(i). Let $\{\xi_i\}_{i=1}^n$ be a sequence of i.i.d. standard normal random variables, drawn independently of the original data. We define the estimated multiplier process (EMP) as:
\begin{align}
	\widehat{\G}^m_{\xi}(\theta) 
	:= \frac{1}{\sqrt{nh_{n,\theta}}}\sum_{i=1}^{n} \xi_i  \frac{(\iota_2 - \iota_3)^\top \widebar{\varGamma}_p^{-1}\bar{r}_p\left(\frac{X_i-x_0}{h_{n,\theta}}\right) K\left(\frac{X_i-x_0}{h_{n,\theta} }\right) \hat{\varepsilon}^m(Y_i,X_i,\theta)}{(b'(x_0^+)-b'(x_0^-)) \hat{f}_X(x_0)}
	\label{eq:def.G.xi.m}
\end{align}
for each $\theta\in\widebar{\varTheta}$. This process requires estimators for the unknown components in the asymptotic representation. The residuals are computed using the fitted values from the constrained regression,
\(\hat{\varepsilon}^m(y,x,\theta):=\left(\varphi(y,\theta) - \bar{r}_p(x-x_0)\hat{\alpha}(\theta)\right)\I\left(\left|x-x_0\right|\leq h_{n,\theta}\right)\), whose uniform consistency is established in Lemma \ref{lem:asy.first.stage}. The density \(f_X(x_0)\) is estimated using a standard kernel density estimator \(\hat{f}_X(x_0)=\frac{1}{nv_n}\sum_{i=1}^{n}K\left((X_i-x_0)/v_n\right)\). 

Approximating the asymptotic distribution for the Type 2 estimator $\widehat{\rkd}_{\psi|m}$ requires to generate valid bootstrap analogues for both. The process $\G^m$ is simulated using the multiplier bootstrap $\widehat{\G}^m_{\xi}$ as described previously. For the quantile component $\G^Q$, we employ the pivotal method developed by \cite{chiang2019causal} and \cite{qu2015nonparametric}. Let $\{U_i\}_{i=1}^n$ be a sequence of i.i.d. Uniform$(0,1)$ random variables, drawn independently of the data. The estimated pivotal process for $\G^Q$ is then constructed as:
\begin{align*}
	\widehat{\G}^Q_{u}(\tau) 
	:= \frac{1}{\sqrt{nh_{n,\tau}}}\sum_{i=1}^{n}\frac{(\iota_2 - \iota_3)^\top \widebar{\varGamma}_p^{-1}\bar{r}_p\left(\frac{X_i-x_0}{h_{n,\tau}}\right) K\left(\frac{X_i-x_0}{h_{n,\tau}}\right) \big(\tau - \I(U_i \leq \tau)\big)}{(b'(x_0^+)-b'(x_0^-)) \hat{f}_X(x_0) \hat{f}_{Y|X}(\widehat{Q}_{Y|X}(\tau|x_0)|x_0)}.
\end{align*}
The bootstrap analogue for the composite limiting process $\G^{\psi|m}$ is then constructed by combining these two simulated processes using the estimated Hadamard derivative of the functional $\psi$:
 \begin{align}
 	\widehat{\G}^{\psi|m}_{\xi,u}(\theta') 
 	 := \hat{\psi}'_{(\rkd_m, \qrkd)}\left( \frac{\widehat{\G}^m_{\xi}(\cdot)}{\sqrt{\varsigma(\cdot)}}, \frac{\widehat{\G}^Q_{u}(\cdot)}{\sqrt{c(\cdot)}}\right)(\theta'). 
 	\label{eq:def.G.psi.m}
 \end{align}
 
The following high-level assumption is required to establish the uniform validity of our resampling procedures.
\begin{assumption}[First Stage Estimation]~\label{A:asy.first.stage.esti}
	\begin{itemize}		
		\item[(i)]
		\(\hat{f}_X(x_0) = f_X(x_0) + o_{P}(1)\), and \(\hat{f}_{Y|X}(y|x_0) = f_{Y|X}(y|x_0) + o_{P}(1)\) uniformly in \(y \in \mathcal{Y}_{x_0}\). 
		
		\item[(ii)]
		For each \((g, h )\in \ell^{\infty}(\widebar{\varTheta}) \times \ell^{\infty}(\mathcal{T}) \), the estimated Hadamard derivative satisfies 
		\[\hat{\psi}'_{(\rkd_m, \qrkd)}(g,h)(\cdot) = \psi'_{(\rkd_m, \qrkd)}(g,h)(\cdot) + o_{P}(1)\] uniformly on \(\widebar{\varTheta}'\)
	\end{itemize}
\end{assumption}
Assumption \ref{A:asy.first.stage.esti} ensures the consistency of the nuisance components used in constructing $\widehat{\G}^m_{\xi}$ and $\widehat{\G}^{\psi|m}_{\xi,u}$. The role of Part (ii) can be illustrated using the Lorenz effect $\Delta_L$ as an example. For this effect, the functional $\psi_L$ is bilinear, and its Hadamard derivative, $[\psi_L]'$, depends on the nuisance parameters $\mu_0$ and the function $Q_{Y|X}(\cdot|x_0)$. In this context, Assumption \ref{A:asy.first.stage.esti}(ii) is a high-level condition requiring uniform consistency of the estimators for these nuisance components, i.e., $\hat{\mu}_0 \pto{P} \mu_0$ and $\sup_{\tau \in \mathcal{T}} |\widehat{Q}_{Y|X}(\tau|x_0) - Q_{Y|X}(\tau|x_0)| \pto{P} 0$.

Let \(\wto{P}{M}\) denote the conditional weak convergence, where the subscript $M$ indicates
conditional expectation over the weights $M$ given the remaining data (\citealp[Page. 20]{kosorok2008introduction}). 
\begin{theorem}[Conditional Weak Convergence]\label{thm:asy.mb.pivotal}
	Suppose Assumptions \ref{A:sharp.kink}--\ref{A:asy.first.stage.esti} hold. Then:
	\begin{itemize}
		\item[(i)]
		\(\widehat{\G}^m_{\xi}(\theta) \wto{P}{\xi} \G^m(\theta)\) uniformly in \(\theta \in \widebar{\varTheta}\).
		
		\item[(ii)]
		\(\widehat{\G}^{\psi|m}_{\xi,u}(\theta') \wto{P}{\xi\times u} \G^{\psi|m}(\theta')\) uniformly in \(\theta' \in \widebar{\varTheta}'\). 
	\end{itemize}
\end{theorem}
Theorem \ref{thm:asy.mb.pivotal} establishes that the empirical distributions of the simulated processes, $\widehat{\G}^m_{\xi}(\theta)$ and $\widehat{\G}^{\psi|m}_{\xi,u}(\theta')$, validly approximate the asymptotic distributions of the normalized processes of $\widehat{\rkd}_m(\theta) $ and $\widehat{\rkd}_{\psi|m}(\theta')$, respectively. This result enables the construction of asymptotically valid uniform confidence bands and hypothesis tests in practice.

\subsection{Hypothesis Tests and Uniform Confidence Bands}
Building on the asymptotic results of Theorems \ref{thm:asy} and \ref{thm:asy.mb.pivotal}, this section details the procedure for conducting uniform inference on the local treatment effects, including uniform hypothesis testing and the construction of uniform confidence bands. Our goal is to conduct inference on the functions $\theta \mapsto \Delta_m(\theta)$ and $\theta' \mapsto \Delta_{\psi|m}(\theta')$, which represent our \textit{Type 1} and \textit{Type 2 LTEs}, respectively. 

To test for the uniform significance of the treatment effect, we consider the null hypothesis that the LTE function is identically zero over the region of interest. For our Type 1 and Type 2 effects, respectively, the hypotheses are:
\begin{align*}
	&\mathscr{H}_{0,m}^S: \Delta_m(\theta) = 0 \quad \text{for all $\theta \in \widebar{\varTheta}$},\\
	\text{and }\;
	&\mathscr{H}_{0,\psi|m}^S: \Delta_{\psi|m}(\theta') = 0 \quad \text{for all $\theta' \in \widebar{\varTheta}'$},
\end{align*}
where \(\widebar{\varTheta} \subset \varTheta\) and \(\widebar{\varTheta}' \subset \varTheta'\) are compact spaces of interests. We test these hypotheses using Kolmogorov–Smirnov type test statistics, which measure the maximum deviation of the estimated process from zero:
\begin{align*}
	&W_{m}^S(\widebar{\varTheta}) 
	= \sup_{\theta \in \widebar{\varTheta}} \left|\sqrt{nh_{n,\theta}^3}\,\widehat{\rkd}_m(\theta)\right|, \\
	\text{and }\; 
	&W_{\psi|m}^S(\widebar{\varTheta}') 
	= \sup_{\theta' \in \widebar{\varTheta}'} \left|\sqrt{nh_n^3}\, \widehat{\rkd}_{\psi|m}(\theta') \right|.
\end{align*}

To test for the homogeneity of the treatment effect, we consider the null hypothesis that the LTE function is constant, though not necessarily zero. The hypotheses are:
\begin{align*}
	&\mathscr{H}_{0,m}^H: \Delta_m(\theta_1) = \Delta_m(\theta_2) \quad \text{for all $\theta_1, \theta_2 \in \widebar{\varTheta}$},\\
	\text{and }\;
	&\mathscr{H}_{0,\psi|m}^H: \Delta_{\psi|m}(\theta'_1) = \Delta_{\psi|m}(\theta'_2) \quad \text{for all $\theta'_1, \theta'_2 \in \widebar{\varTheta}'$},
\end{align*}
These hypotheses can be tested by measuring the deviation from the mean over the region of interest. The specific statistics are defined as:
\begin{align*}
	&W_{m}^H(\widebar{\varTheta}) 
	= \sup_{\theta \in \widebar{\varTheta}} \left| \sqrt{nh_{n,\theta}^3} \left(\widehat{\rkd}_m(\theta) - \frac{1}{|\widebar{\varTheta}|} \int_{\widebar{\varTheta}} \widehat{\rkd}_m(\theta)\,d\theta \right)  \right|, \\
	\text{and }\;
	&W_{\psi|m}^H(\widebar{\varTheta}')
	= \sup_{\theta' \in \widebar{\varTheta}'} \left| \sqrt{nh_n^3} \left(\widehat{\rkd}_{\psi|m}(\theta') - \frac{1}{|\widebar{\varTheta}'|} \int_{\widebar{\varTheta}'} \widehat{\rkd}_{\psi|m}(\theta')\,d\theta'  \right) \right|.
\end{align*}

The following corollary, which is a direct consequence of Theorem \ref{thm:asy}, establishes the asymptotic distribution of our test statistics under their respective null hypotheses.
\begin{corollary}\label{cory:asy.test}
	Suppose Assumptions \ref{A:sharp.kink}--\ref{A:asy.first.stage.esti} hold. Then,
	\begin{itemize}
		\item[(i)]
		\(W_{m}^S(\widebar{\varTheta}) \leadsto \sup_{\theta \in \widebar{\varTheta}}\left|\G^m(\theta)\right|\) under the null hypothesis \(\mathscr{H}_{0,m}^S\); 
		
		\(W_{m}^H(\widebar{\varTheta}') \leadsto \sup_{\theta \in \widebar{\varTheta} } \left|\G^m(\theta) - \frac{1}{|\widebar{\varTheta}|} \int_{\widebar{\varTheta}} \G^m(\theta) \,d\theta  \right|\) under the null hypothesis \(\mathscr{H}_{0,m}^H\).

		\item[(ii)]
		\(W_{\psi|m}^S(\widebar{\varTheta}) \leadsto \sup_{\theta \in \widebar{\varTheta}}\left|\G^{\psi|m}(\theta)\right|\) under the null hypothesis \(\mathscr{H}_{0,\psi|m}^S\); 
		
		\(W_{\psi|m}^H(\widebar{\varTheta}')\leadsto \sup_{\theta' \in \widebar{\varTheta}'} \left|\G^{\psi|m}(\theta') - \frac{1}{|\widebar{\varTheta}'|} \int_{\widebar{\varTheta}'} \G^{\psi|m}(\theta') \,d\theta'  \right|\) under the null hypothesis \(\mathscr{H}_{0,\psi|m}^H\).
		
	\end{itemize}
\end{corollary}
In practice, the asymptotic distributions of the test statistics in Corollary \ref{cory:asy.test} can be simulated based on the estimated limiting processes \(\widehat{G}^m_{\xi}\) and $\widehat{G}^{\psi|m}_{\xi,u}$, respectively. 


Building on the bootstrap validity established in Theorem \ref{thm:asy.mb.pivotal}, we now detail the construction of uniform confidence bands for our LTEs. The procedure for a generic LTE function, $\Delta(\cdot)$, follows a standard two-step process:

\textit{Step 1: Estimate the Critical Value.} We simulate the distribution of the supremum norm of the relevant bootstrap process. For $\lambda\in (0,1)$, the estimated $(1-\lambda)$ critical value, $\hat{c}(1-\lambda)$, is then obtained as the $(1-\lambda)$ sample quantile of this simulated distribution.

\textit{Step 2: Construct the Band.} The $100(1-\lambda)\%$ uniform confidence band for the function $\Delta(\cdot)$ is then formed by:
$$ \left[ \widehat{\Delta}(\cdot) \pm \hat{c}(1-\lambda)\big/\sqrt{nh_{n,\cdot}^3}\right], $$
where $\widehat{\Delta}(\cdot)$ is the corresponding point estimator $\widehat{\rkd}_m(\theta)$ or $\widehat{\rkd}_{\psi|m}(\theta')$. Specifically, for the Type 1 LTE $\Delta_m$, we use the estimated bootstrap process $\widehat{\G}^m_{\xi}$ to compute $\hat{c}_m(1-\lambda)$, and for the Type 2 LTE $\Delta_{\psi|m}$, we use the estimated composite process $\widehat{\G}^{\psi|m}_{\xi,u}$ to compute $\hat{c}_{\psi|m}(1-\lambda)$.

This section concludes by detailing the implementation of our inference procedures for the specific LTEs discussed previously. Bandwidth selection procedures for each effect are provided in Appendix \ref{sec:apdx.bandwidth}. Let $B$ be the number of bootstrap replications.
\setcounter{example}{0}
\begin{example}[Average Effect, Continued]
	The LATE \(\Delta_\mu\) is estimated by $\widehat{\rkd}_\mu$.  For each replication $b = 1, \dots, B$, we generate an i.i.d. sequence of \(N(0,1)\) random weights $\{\xi_i^b\}_{i=1}^n$, drawn independently of the data.  Then, the distribution of \(\sqrt{nh_n^3}(\widehat{\rkd}_{\mu}-\Delta_\mu)\) is approximated by the empirical distribution of \(\{\widehat{\G}^{\mu}_{\xi^b}\}_{b=1}^B\), where
	\begin{align*}
		\widehat{\G}^{\mu}_{\xi^b}
		:=\sum_{i=1}^{n} \xi^b_i  \frac{(\iota_2 - \iota_3)^\top \widebar{\varGamma}_p^{-1}\bar{r}_p\left(\frac{X_i-x_0}{h_n}\right) K\left(\frac{X_i-x_0}{h_n }\right) \hat{\varepsilon}^\mu(Y_i,X_i)}{(b'(x_0^+)-b'(x_0^-)) \sqrt{nh_n} \hat{f}_X(x_0)}.
	\end{align*}
	The random error is estimated by \(\hat{\varepsilon}^\mu(y,x)=\left(y- \bar{r}_p(x-x_0)\widehat{\alpha}\right)\I\left(\left|x-x_0\right|\leq h_n\right)\). 
	To test the significance hypothesis, $\Delta_\mu = 0$, we first compute the test statistic $W_\mu^S = |\sqrt{nh_n^3}\,\widehat{\rkd}_\mu|$.  The $(1-\lambda)$ critical value, $\hat{c}_\mu(1-\lambda)$, is the $(1-\lambda)$ sample quantile of the simulated distribution $\{|\widehat{\G}^\mu_{\xi^b}|\}_{b=1}^B$. We reject the null if $W_\mu^S > \hat{c}_\mu(1-\lambda)$. The bootstrap $p$-value is calculated as $\hat{p}^S_\mu = \frac{1}{B}\sum_{b=1}^{B}\I(|\widehat{\G}^\mu_{\xi^b}| > W_\mu^S)$. Finally, the uniform confidence band is constructed as $\left[\widehat{\rkd}_\mu \pm \hat{c}_\mu(1-\lambda)/\sqrt{nh_n^3}\right]$. 
\end{example}

\begin{example}[Distributional and Quantile Effects, Continued]~\\
	The LDTE $\Delta_{Id}(\cdot)$ is estimated by  $\widehat{\rkd}_F(\cdot)$.
	A common application is to test properties of the LDTE $\Delta_{Id}$ over a range of quantiles, i.e., to conduct inference on the process $\{\Delta_{Id}(y_\tau)\}_{\tau \in \mathcal{T}}$.
	For each replication $b = 1, \dots, B$, generate an i.i.d. sequence of \(N(0,1)\) weights $\{\xi_i^b\}_{i=1}^n$.
	We construct the $b$th draw of the bootstrap process $\widehat{\G}^{Id}_{\xi^b}(\cdot)$ as 
	\begin{align*}
		\widehat{\G}^{Id}_{\xi^b}(\theta):=\sum_{i=1}^{n} \xi^b_i  \frac{(\iota_2 - \iota_3)^\top \widebar{\varGamma}_p^{-1}\bar{r}_p\left(\frac{X_i-x_0}{h_{n,\theta}}\right) K\left(\frac{X_i-x_0}{h_{n,\theta} }\right) \hat{\varepsilon}^F(Y_i,X_i,\theta)}{(b'(x_0^+)-b'(x_0^-)) \sqrt{nh_{n,\theta}} \hat{f}_X(x_0)},
	\end{align*}
	where the residual is computed as $\hat{\varepsilon}^F(y,x,\theta)=\left(\I(y\leq \theta) - \bar{r}_p(x-x_0)\widehat{\alpha}\right)\I\left(\left|x-x_0\right|\leq h_n\right)$. 
	For each replication $b$, evaluate the bootstrap process $\widehat{\G}^{Id}_{\xi^b}(\cdot)$ at the estimated quantile points $\{\hat{y}_\tau\}_{\tau \in \mathcal{T}}$ to obtain the simulated process $\{\widehat{\G}^{Id}_{\xi^b}(\hat{y}_\tau)\}_{\tau \in \mathcal{T}}$.

	To test the hypothesis of uniform treatment significance $\Delta_{Id}(y_\tau) = 0$ for all $\tau \in \mathcal{T}$, we first compute  the test statistic \(W_{Id}^S:=\sup_{\tau \in \mathcal{T}}|\sqrt{nh_{n,y_\tau}^3}\, \widehat{\rkd}_F(\hat{y}_\tau)|\). Then, we use $\{\widehat{\G}^S_{Id,b}:=\sup_{\tau\in\mathcal{T}}|\widehat{\G}^{Id}_{\xi^b}(\hat{y}_\tau)|\}_{b=1}^B$ 
	to approximate the distribution of \(W_{Id}^S\). The $(1-\lambda)$ critical value, $\hat{c}_{Id}(1-\lambda)$, is the $(1-\lambda)$ sample quantile of $\{\widehat{\G}^S_{Id,b}\}_{b=1}^B$. The bootstrap $p$-value is given by $\hat{p}^S_{Id}:=\frac{1}{B}\sum_{b=1}^{B}\I(\widehat{\G}^S_{Id,b}>W_{Id}^S)$.
	
	To test the hypothesis of uniform treatment homogeneity $\Delta_{Id}(y_{\tau_1}) = \Delta_{Id}(y_{\tau_2})$ for all $\tau_1, \tau_2 \in \mathcal{T}$, we first compute the test statistic 
	\[W^H_{Id}:=\sup_{\tau\in\mathcal{T}} \left| \sqrt{nh_{n,y_\tau}^3}\left( \widehat{\rkd}_F(\hat{y}_\tau) - \frac{1}{|\mathcal{T}|} \int_{\mathcal{T}} \widehat{\rkd}_F(\hat{y}_\tau)\,d\tau  \right) \right|. 
	\]
	Then, we use $\{\widehat{\G}^H_{Id,b}:=\sup_{\tau\in\mathcal{T}}|\widehat{\G}^{Id}_{\xi^b}(\hat{y}_\tau) - \frac{1}{|\mathcal{T}|} \int_{\mathcal{T}} \widehat{\G}^{Id}_{\xi^b}(\hat{y}_\tau)\,d\tau|\}_{b=1}^B$ 
	to approximate the distribution of $W^H_{Id}$. The critical value and $p$-value are calculated analogously to the significance test. Finally, the uniform confidence band is constructed as $$\left[\widehat{\rkd}_F(\hat{y}_\tau) \pm \hat{c}_{Id}(1-\lambda)/\sqrt{nh_{n,y_\tau}^3}\right].$$
	
	Inference for the LQTE, $\Delta_Q(\cdot)$, which is estimated by $\widehat{\qrkd}(\cdot)$, is conducted using the pivotal method detailed in Section \ref{sec:esti.infer}. This involves generating i.i.d. Uniform$(0,1)$ random variables to construct the pivotal process $\{\widehat{\G}^Q_{u^b}(\tau)\}_{\tau \in \mathcal{T}}$ for each replication $b$.
	The subsequent steps for constructing test statistics and obtaining critical values are analogous to those for the LDTE. For the sake of brevity, we omit the detailed formulas and refer readers to the comprehensive procedure outlined in Appendix C.2 of \cite{chiang2019causal}, which our method follows directly.
\end{example}

\begin{example}[Lorenz Effect, Continued]
	By Equation (\ref{eq:esti.llte}), the LLTE \(\Delta_{L}(\tau)\) is estimated by \(\widehat{\rkd}_{\psi_L|\mu}(\tau)\).
	For each bootstrap iteration \(b\in\{1,\ldots,B\}\), we draw the i.i.d. sequence \(\{\xi^b_i\}_{i=1}^n\) from \(N(0,1)\), and the i.i.d. sequence \(\{u^b_i\}_{i=1}^n\) from Uniform$(0,1)$, respectively.
	Then, we approximate the distribution of \(\sqrt{n(h^L_{n,\tau})^3}(\widehat{\rkd}_{\psi_L|\mu}(\tau)-\Delta_{L}(\tau))\) by the empirical distribution of \(\{\widehat{\G}^L_{\xi^b,u^b}(\tau)\}_{b=1}^B\) for each \(\tau \in \mathcal{T}\), where 
	\begin{align*}
		\widehat{\G}^L_{\xi^b,u^b}(\tau) 
		:= \frac{1}{\hat{\mu}_0} \left(\int_{0}^{\tau} \widehat{\G}^Q_{u^b}(u)\,du - \widehat{L}_{Y|X}(\tau|x_0) \cdot \widehat{\G}^{\mu}_{\xi^b} \right).
	\end{align*}
	It is important to note the bandwidth choices used in the bootstrap procedures. For the mean effect bootstrap, $\widehat{\G}^{\mu}_{\xi^b}$, the single baseline bandwidth $h_n$ is employed. In contrast, for the quantile effect bootstrap, $\widehat{\G}^Q_{u}(\tau)$, the pointwise bandwidth $h_{n,\tau}$ is used. As established by Lemma \ref{lem:asy.qrkd.cs19}, using the pointwise bandwidth $h_{n,\tau}$ ensures that this bootstrap process correctly targets the desired limiting process $\G^Q(\tau)$, rather than the scaled version. 

	To test the hypothesis of uniform treatment significance $\Delta_L(\tau) = 0$ for all \(\tau \in \mathcal{T}\), we first compute the test statistic $W_{L}^S:=\sup_{\tau \in \mathcal{T}}|\sqrt{n(h^L_{n,\tau})^3}\, \widehat{\rkd}_{\psi_L|\mu}(\tau)|$. Then, we use 
	$\{\widehat{\G}^S_{L,b}:=\sup_{\tau\in\mathcal{T}}|\widehat{\G}^L_{\xi^b,u^b}(\tau)|\}_{b=1}^B$ 
	to approximate the distribution of $W_{L}^S$.  The \((1-\lambda)\)th critical value, $\hat{c}_{L}(1-\lambda)$, is \((1-\lambda)\) sample quantile of $\{\widehat{\G}^S_{L,b}\}_{b=1}^B$. The $p$-value is given by \(\hat{p}^S_{L}:=\frac{1}{B}\sum_{b=1}^{B}\I(\widehat{\G}^S_{L,b}>W_{L}^S)\).
	
	To test the hypothesis of uniform treatment homogeneity $\Delta_{L}(\tau_1) = \Delta_{L}(\tau_2)$ for all $\tau_1, \tau_2 \in \mathcal{T}$, we first compute the test statistic the test statistic 
	\[W^H_{L}:=\sup_{\tau\in\mathcal{T}} \left| \sqrt{n(h^L_{n,\tau})^3} \left(\widehat{\rkd}_{\psi_L|\mu}(\tau) - \frac{1}{|\mathcal{T}|} \int_{\mathcal{T}} \widehat{\rkd}_{\psi_L|\mu}(\tau)\,d\tau  \right)  \right|.
	\]
	Then, we use $\{\widehat{\G}^H_{L,b}:=\sup_{\tau\in\mathcal{T}}|\widehat{\G}^L_{\xi^b,u^b}(\tau) - \frac{1}{|\mathcal{T}|} \int_{\mathcal{T}} \widehat{\G}^L_{\xi^b,u^b}(\tau)\,d\tau|\}_{b=1}^B$ 
	to approximate the distribution of $W^H_{L}$. The critical value is \((1-\lambda)\) sample quantile of $\{\widehat{\G}^H_{L,b}\}_{b=1}^B$. The $p$-value of the treatment homogeneity test is given by \(\hat{p}^H_{L}=\frac{1}{B}\sum_{b=1}^{B}\I(\widehat{\G}^H_{L,b} > W^H_{L})\). Finally, the uniform confidence band is constructed as $$\left[\widehat{\rkd}_{\psi_L|\mu}(\tau) \pm \hat{c}_{L}(1-\lambda)/\sqrt{n(h^L_{n,\tau})^3}\right].$$

\end{example}

\section{Numerical Illustrations}\label{sec:numerical.illustration}
\subsection{Empirical Application} \label{sec:emiprical}
In this section, we apply our framework to a classic dataset from the Continuous Wage and Benefit History (CWBH) to analyze the causal effects of unemployment insurance (UI) benefits on unemployment durations. The data cover unemployed individuals in Louisiana over two periods: September 1981--September 1982 and September 1982--December 1983. Our analysis builds upon the work of \cite{landais2015assessing}, to which we refer readers for detailed descriptive statistics and institutional background.
The dataset contains the following variables: (i) Running variable ($X$): The highest quarterly wage in an individual's base period. (ii) Treatment ($B$): The weekly UI benefit amount received. (iii) Outcome ($Y$): The duration of unemployment, measured in two ways: (a) the number of weeks benefits were claimed (UI Claimed), and (b) the number of weeks benefits were paid (UI Paid).
This setting is a canonical example of a sharp regression kink design. The UI benefit amount $B$ is a deterministic, piecewise-linear function of the base wage $X$: \(b(x)=\min\{0.04\cdot x, b_{max}\}\). The maximum benefit level $b_{max}$, which defines the location of the kink point $x_0$, changed between the two periods. For the 1981--1982 period, $b_{max}$ was $\$ 4,575$, whereas for the 1982--1983 period, it increased to $\$ 5,125$.

\begin{figure}[h]
	\caption{Estimates of Local Distributional and Lorenz Treatment Effects, September 1981 -- September 1982}\label{fig:empirical.1981}
	\centering
	\makebox[\linewidth]{\includegraphics[width=\textwidth]{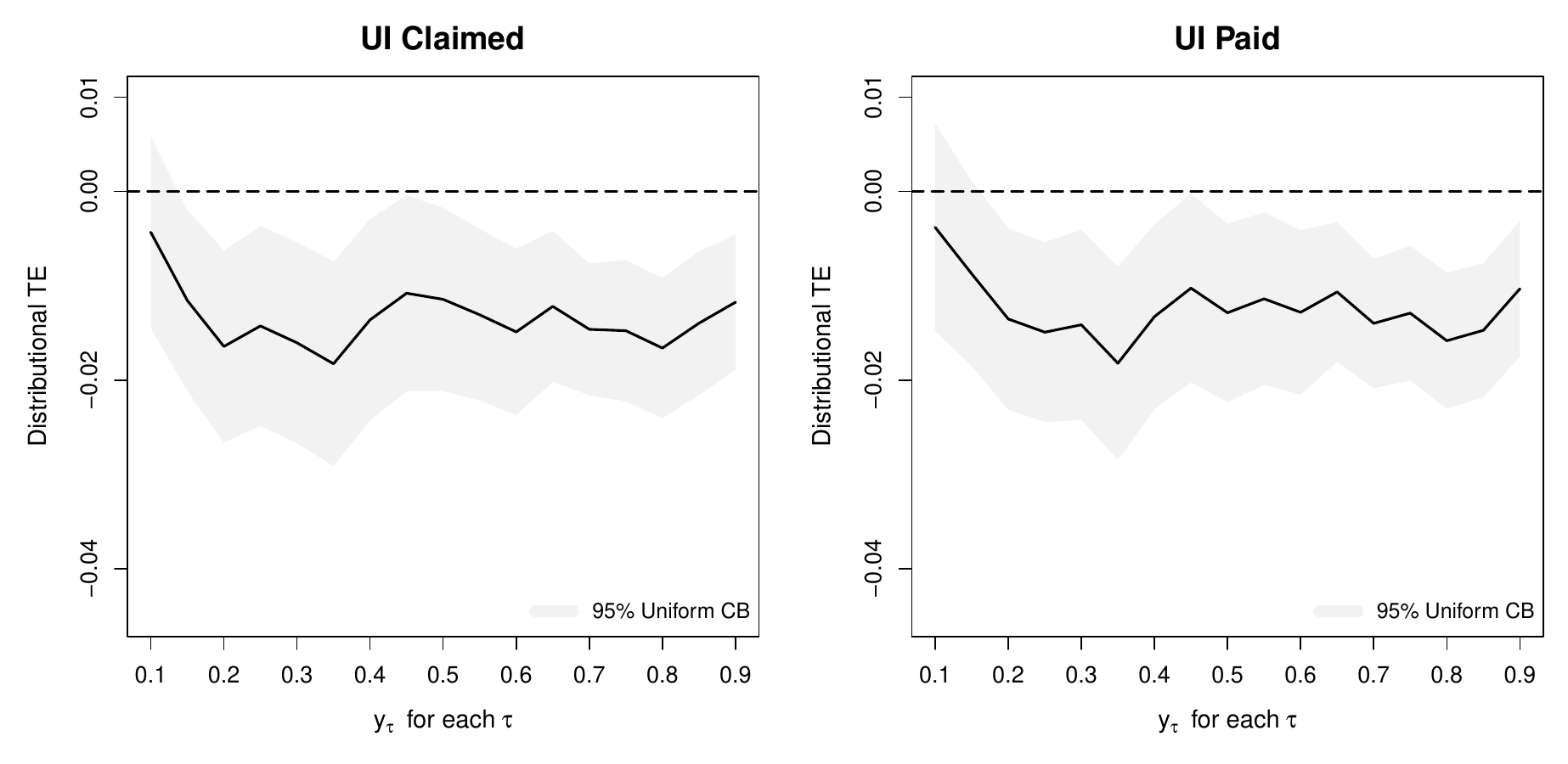}}
	\makebox[\linewidth]{\includegraphics[width=\textwidth]{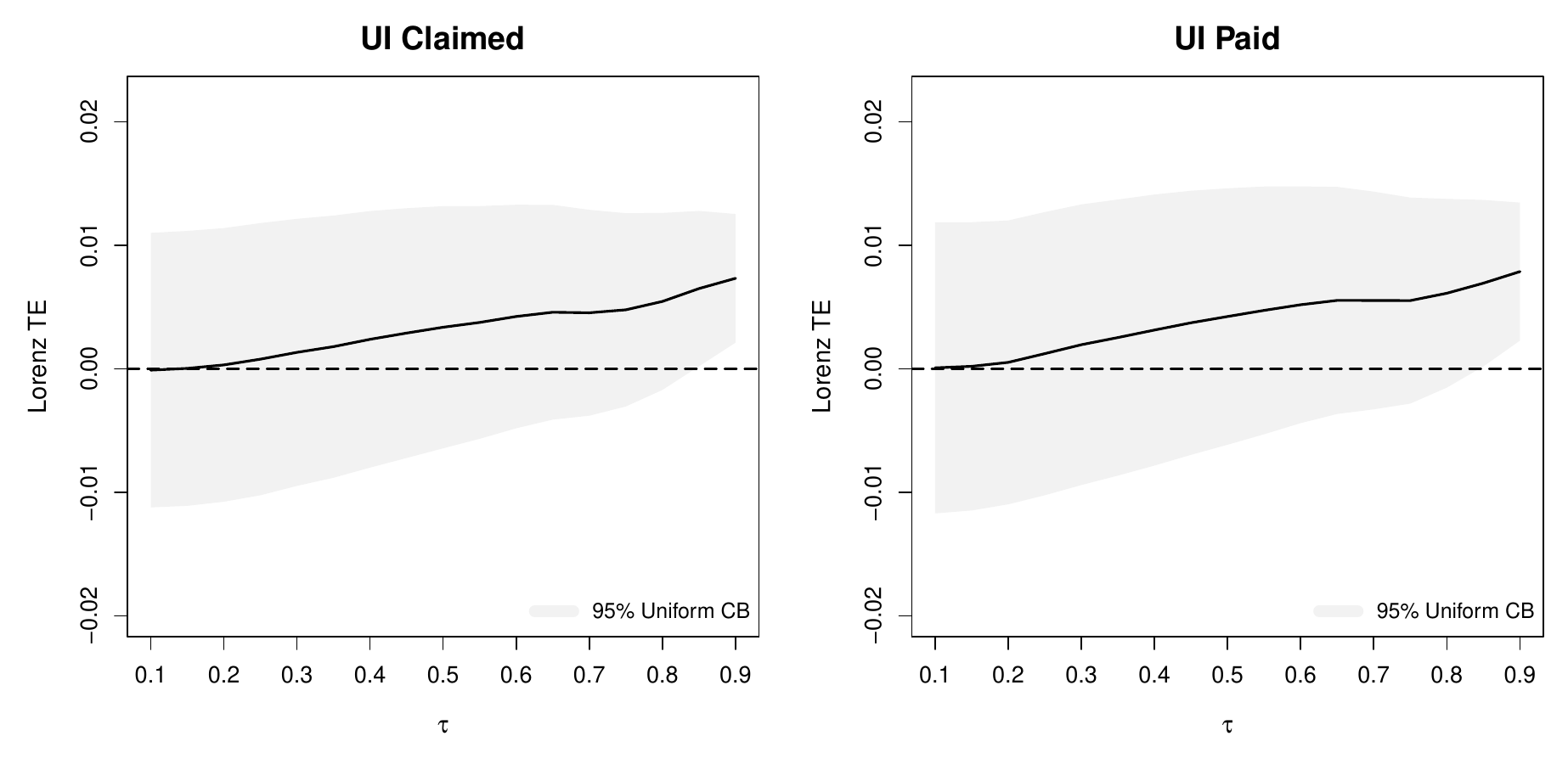}}
\end{figure}

\begin{figure}[h]
	\caption{Estimates of Local Distributional and Lorenz Treatment Effects, September 1982 -- December 1983}\label{fig:empirical.1982}
	\centering
	\makebox[\linewidth]{\includegraphics[width=\textwidth]{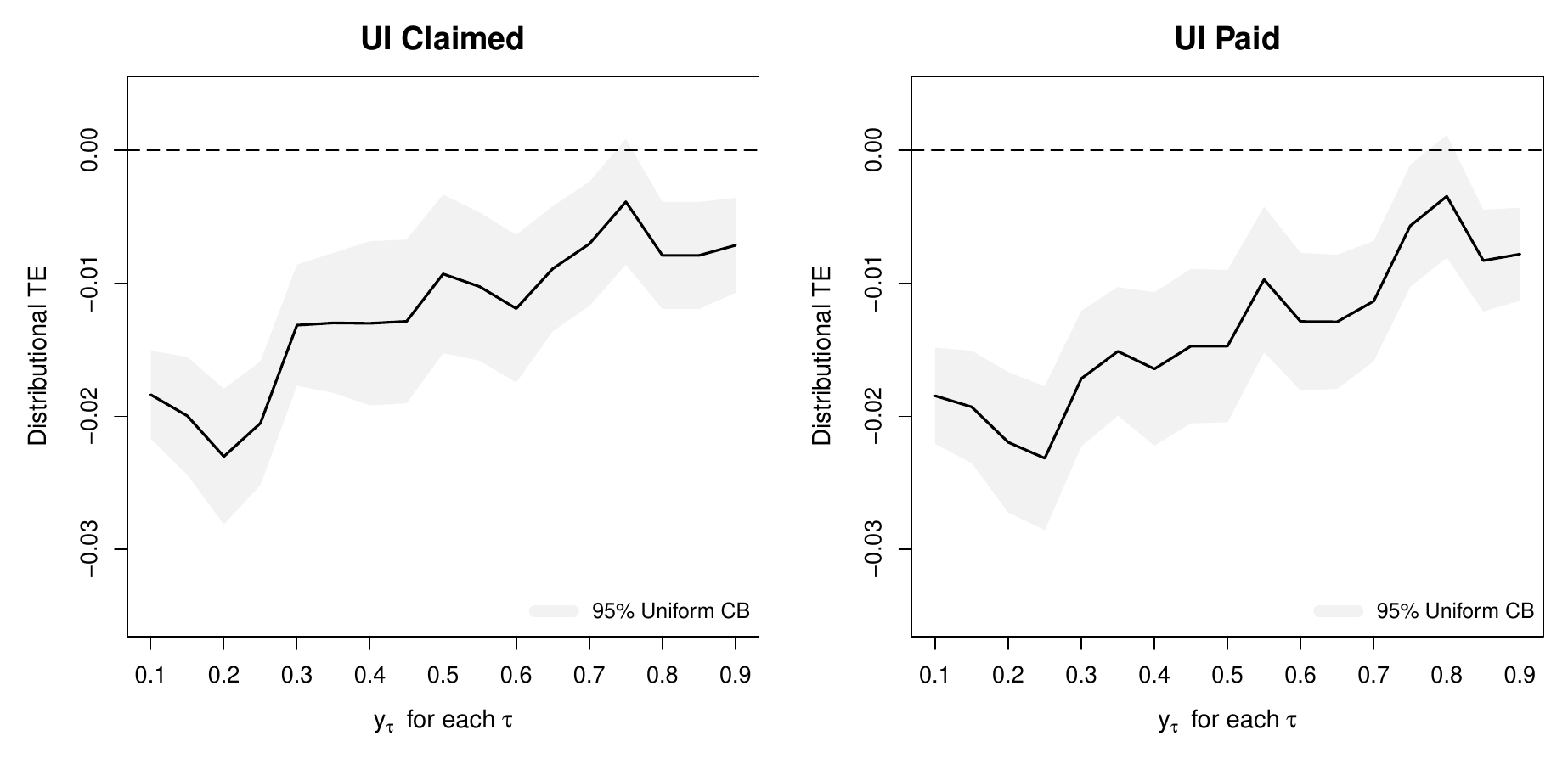}}
	\makebox[\linewidth]{\includegraphics[width=\textwidth]{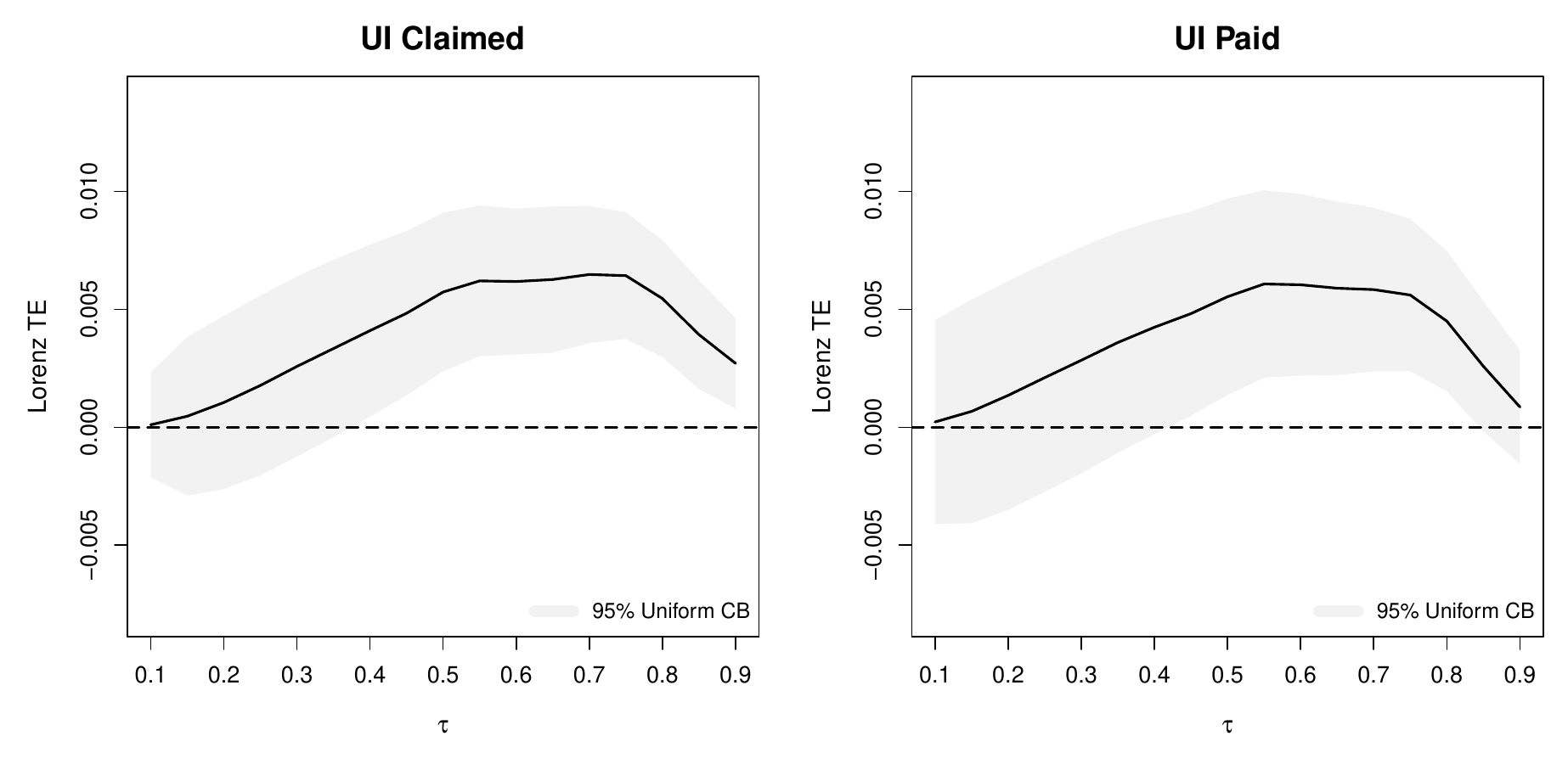}}
\end{figure}

Building on the theoretical framework, this analysis examines the local distributional (LDTE) and local Lorenz (LLTE) treatment effects of UI benefits on unemployment durations. This focus extends the existing literature, which has primarily examined the average effect  (\citealp{landais2015assessing})  and the quantile effect  (\citealp{chiang2019causal}). We estimate the LDTE, $\Delta_{Id}(y_\tau)$, and the LLTE, $\Delta_L(\tau)$, over a grid of quantiles $\tau$. The full estimated effect functions for both time periods are displayed graphically in Figures \ref{fig:empirical.1981} and \ref{fig:empirical.1982}. For specific numerical results, Tables \ref{tb:empirical.1981} and \ref{tb:empirical.1982} present point estimates and standard errors for these effects at selected quantiles $\tau \in \{0.1, 0.2, \dots, 0.9\}$, along with $p$-values from the corresponding uniform hypothesis tests. All estimation and inference procedures follow the methods detailed in Section \ref{sec:esti.infer}. Consistent with \cite{chiang2019causal}, we employ a tricube kernel function, $K(u) = \frac{70}{81}(1-|u|^3)^3\I(|u|\leq 1)$, and set the local polynomial order to $p=2$ for bias-correction.  Selection of the MSE-optimal bandwidths follows the procedures detailed in Algorithms \ref{alg:bandwidth.m} and \ref{alg:bandwidth.lorenz} in Appendix \ref{sec:apdx.bandwidth}.

Panel A of the tables presents the results for the LDTE. The findings are summarized below.
First, the estimated LDTE is negative across both time periods examined. This implies that a marginal increase in UI benefits reduces the cumulative probability of an unemployment spell ending within any given number of weeks for individuals at the kink wage.
Second, the magnitude of the estimated effect varies with duration and changes between periods. In the 1981--1982 period, the effect's magnitude is smaller at shorter durations (e.g., the 10th percentile) and increases at longer durations. This pattern appears reversed in the 1982--1983 period, where the effect is largest in magnitude at short durations (e.g., the 20th percentile) and smaller at longer durations.
Third, the LDTE is statistically significant at most quantile levels in both periods. Furthermore, formal tests for homogeneity are strongly rejected, confirming that the distributional effect is not constant. The results are also robust across the two different measures of unemployment duration (UI Claimed vs. UI Paid).

Panel B presents the results for the LLTE, which measures the effect of UI benefits on the inequality of the duration distribution. The estimated LLTE is positive at all quantiles considered in both periods. This indicates that a marginal increase in UI benefits increases the dispersion (inequality) of unemployment durations for individuals at the kink wage.
This finding is consistent with the quantile effect patterns reported in \cite{chiang2019causal}. Their observation---that benefit increases have a larger effect on longer unemployment spells---implies an increase in dispersion, which the LLTE directly measures.
Similar to the LDTE, the estimated Lorenz effects are statistically significant in both periods. Tests also reject the null hypothesis of homogeneity. The findings remain consistent across the two measures of unemployment duration.

\begin{table}[h]
	\centering
	\fontsize{10}{15}\selectfont
	\begin{threeparttable}
		\caption{Estimation and Inference of the Local Distributional and Lorenz Treatment Effects, September 1981 -- September 1982}\label{tb:empirical.1981}
		\begin{tabular}{l@{\hspace{3em}} l@{\hspace{3em}} r@{\hspace{1.5em}} r@{\hspace{3em}} r@{\hspace{1.5em}} r}
			\hline	
			\multicolumn{6}{l}{September 1981--September 1982} \\
			Dependent Variable & & \multicolumn{2}{l}{Ui Claimed } &  \multicolumn{2}{l}{Ui Paid} \\
			\hline
			Panel A. \; LDTE \(\Delta_{Id}(y)\) 
			& \(y=y_{0.1}\) & $-$0.004 & (0.003) & $-$0.004 & (0.003) \\
			& \(y=y_{0.2}\) & $-$0.016 & (0.004) & $-$0.014 & (0.003) \\
			& \(y=y_{0.3}\) & $-$0.016 & (0.004) & $-$0.014 & (0.004) \\
			& \(y=y_{0.4}\) & $-$0.014 & (0.004) & $-$0.013 & (0.004) \\
			& \(y=y_{0.5}\) & $-$0.011 & (0.004) & $-$0.013 & (0.004) \\
			& \(y=y_{0.6}\) & $-$0.015 & (0.003) & $-$0.013 & (0.003) \\
			& \(y=y_{0.7}\) & $-$0.015 & (0.002) & $-$0.014 & (0.002) \\
			& \(y=y_{0.8}\) & $-$0.017 & (0.002) & $-$0.016 & (0.002) \\
			& \(y=y_{0.9}\) & $-$0.012 & (0.002) & $-$0.010 & (0.002) \\
			\hline
			Test of significance & $p$-Value & 0.000 & & 0.000 & \\
			Test of Homogeneity & $p$-Value & 0.010  &  & 0.001 & \\
			\hline	
			Panel B. \; LLTE \(\Delta_L(\tau)\)  
			& \(\tau=0.10\) & 0.000 & (0.000) & 0.000 & (0.000) \\
			& \(\tau=0.20\) & 0.000 & (0.000) & 0.001 & (0.000) \\
			& \(\tau=0.30\) & 0.001 & (0.001) & 0.002 & (0.001) \\
			& \(\tau=0.40\) & 0.002 & (0.001) & 0.003 & (0.001) \\
			& \(\tau=0.50\) & 0.003 & (0.002) & 0.004 & (0.002) \\
			& \(\tau=0.60\) & 0.004 & (0.002) & 0.005 & (0.002) \\
			& \(\tau=0.70\) & 0.005 & (0.002) & 0.006 & (0.003) \\
			& \(\tau=0.80\) & 0.005 & (0.003) & 0.006 & (0.003) \\
			& \(\tau=0.90\) & 0.007 & (0.002) & 0.008 & (0.002) \\
			\hline
			Test of significance & $p$-Value & 0.002 & & 0.001 & \\
			Test of Homogeneity & $p$-Value & 0.010 & & 0.018 & \\
			\hline	
		\end{tabular}
		\begin{tablenotes}\footnotesize
			\item \emph{Note}: The numbers in parentheses denote the estimated standard errors.
		\end{tablenotes}
	\end{threeparttable}
\end{table}

\begin{table}[h]
	\centering
	\fontsize{10}{15}\selectfont
	\begin{threeparttable}
		\caption{Estimation and Inference of the Local Distributional and Lorenz Treatment Effects, September 1982 -- December 1983}\label{tb:empirical.1982}
		\begin{tabular}{l@{\hspace{3em}} l@{\hspace{3em}} r@{\hspace{1.5em}} r@{\hspace{3em}} r@{\hspace{1.5em}} r}
			\hline	
			\multicolumn{6}{l}{September 1982--December 1983} \\
			Dependent Variable & & \multicolumn{2}{l}{Ui Claimed } &  \multicolumn{2}{l}{Ui Paid} \\
			\hline
			Panel A. \; LDTE \(\Delta_{Id}(y)\) 
			& \(y=y_{0.1}\) & $-$0.018 & (0.001) & $-$0.018 & (0.001) \\
			& \(y=y_{0.2}\) & $-$0.023 & (0.002) & $-$0.022 & (0.002) \\
			& \(y=y_{0.3}\) & $-$0.013 & (0.002) & $-$0.017 & (0.002) \\
			& \(y=y_{0.4}\) & $-$0.013 & (0.003) & $-$0.016 & (0.002) \\
			& \(y=y_{0.5}\) & $-$0.009 & (0.002) & $-$0.015 & (0.002) \\
			& \(y=y_{0.6}\) & $-$0.012 & (0.002) & $-$0.013 & (0.002) \\
			& \(y=y_{0.7}\) & $-$0.007 & (0.002) & $-$0.011 & (0.002) \\
			& \(y=y_{0.8}\) & $-$0.008 & (0.001) & $-$0.003 & (0.001) \\
			& \(y=y_{0.9}\) & $-$0.007 & (0.001) & $-$0.008 & (0.001) \\
			\hline
			Test of significance & $p$-Value & 0.000 & & 0.000 & \\
			Test of Homogeneity & $p$-Value & 0.000 & & 0.000 & \\
			\hline	
			Panel B. \; LLTE \(\Delta_L(\tau)\)  
			& \(\tau=0.10\) & 0.000 & (0.000) & 0.000 & (0.000) \\
			& \(\tau=0.20\) & 0.001 & (0.000) & 0.001 & (0.000) \\
			& \(\tau=0.30\) & 0.003 & (0.000) & 0.003 & (0.001) \\
			& \(\tau=0.40\) & 0.004 & (0.001) & 0.004 & (0.001) \\
			& \(\tau=0.50\) & 0.006 & (0.001) & 0.006 & (0.001) \\
			& \(\tau=0.60\) & 0.006 & (0.001) & 0.006 & (0.001) \\
			& \(\tau=0.70\) & 0.006 & (0.001) & 0.006 & (0.001) \\
			& \(\tau=0.80\) & 0.005 & (0.001) & 0.005 & (0.001) \\
			& \(\tau=0.90\) & 0.003 & (0.000) & 0.001 & (0.000) \\
			\hline
			Test of significance & $p$-Value & 0.000 & & 0.001 & \\
			Test of Homogeneity & $p$-Value & 0.000 & & 0.000 & \\
			\hline	
		\end{tabular}
				\begin{tablenotes}\footnotesize
			\item \emph{Note}: The numbers in parentheses denote the estimated standard errors.
		\end{tablenotes}
	\end{threeparttable}
\end{table}

\subsection{Simulation Experiment} \label{sec:simulation}
This section evaluates the finite-sample performance of the proposed estimation and inference methods using Monte Carlo simulations. We generate i.i.d. samples $\{(Y_i,B_i,X_i)\}_{i=1}^n$ from the following data generating process:
\begin{align*}
	Y_i &= 1 + 0.5 B_i + X_i + 0.1X_i^2 + 1.5 B_i X_i + (1+2B_i)\varepsilon_i \\
	B_i &= X_i \left(2\cdot\I(X_i \geq 0) - 1\right), \\
	\begin{pmatrix}
		X_i \\
		\varepsilon_i
	\end{pmatrix}
	&\sim N\left(
	\begin{bmatrix}
		0 \\
		0
	\end{bmatrix},
	\begin{bmatrix}
		\sigma_x^2 & \rho \cdot \sigma_x \cdot \sigma_\varepsilon \\
		\rho \cdot \sigma_x \cdot \sigma_\varepsilon  & \sigma_\varepsilon^2
	\end{bmatrix} 
	\right).
\end{align*}
The structure is similar to that in \cite{chiang2019causal} but is extended to include conditional heteroskedasticity through the $(1+2B_i)\varepsilon_i$ term. The kink point is normalized to $x_0=0$. The distributional parameters are chosen to match those in \cite{calonico2014robust}, with $\sigma_x = 0.1781742$, $\sigma_\varepsilon = 0.1295$, and correlation $\rho = 0.25$.

We evaluate the finite-sample performance of estimators for four local treatment effects: the average effect ($\Delta_\mu$), the distributional effect ($\Delta_{Id}$), the quantile effect ($\Delta_Q$), and the Lorenz effect ($\Delta_L$). The LATE $\Delta_\mu$ is estimated at a single point. The LQTE $\Delta_Q(\tau)$ and LLTE $\Delta_L(\tau)$ are evaluated over the quantile grid $\mathcal{T}=\{0.1, 0.2, \dots, 0.9\}$. The LDTE $\Delta_{Id}(y)$ is evaluated at points corresponding to the quantiles in $\mathcal{T}$, i.e., at $y = Q_{Y|X}(\tau|x_0)$ for $\tau \in \mathcal{T}$.
Estimation and inference follow the procedures detailed in Section \ref{sec:esti.infer}. For the LLTE estimator, required numerical integrals are computed using a finer grid, $\mathcal{U}=\{0.01, \dots, 0.99\}$. Nuisance parameters (e.g., $\mu_0$, $Q_{Y|X}$) are estimated as part of the main procedures.
We consider sample sizes \(n\in\{500,1000,2000,4000\}\) and set the number of Monte Carlo replications to 5,000 for each simulation design. In each replication, we employ both the multiplier bootstrap and the pivotal method, each based on 2,500 bootstrap draws, to construct the uniform confidence bands.

\begin{figure}[h] 
	\caption{Simulated Estimates of The Local Treatment Effects across Sample Sizes}\label{fig:simu}
	\centering
	\makebox[\linewidth]{\includegraphics[width=\textwidth]{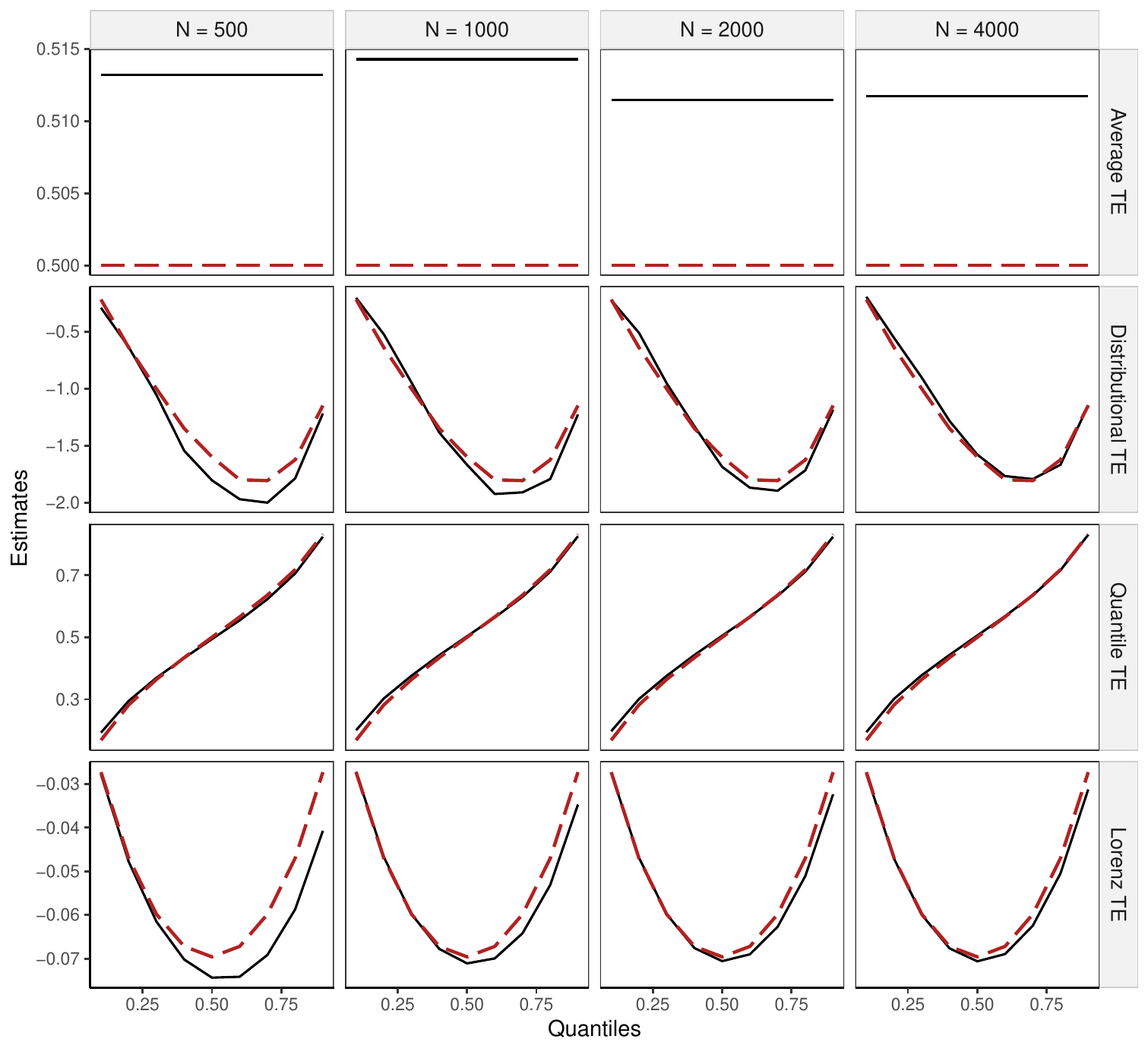}}
	\caption*{\footnotesize 
		\emph{Note:} The red dashed lines correspond to the true values of each local treatment effect, whereas the black solid lines represent their estimates.}
\end{figure}

Figure \ref{fig:simu} plots the estimated effects across simulations against the true parameter values for different sample sizes. Table \ref{tb:simu} reports quantitative measures of finite-sample performance, including the absolute bias ratio ($|\widehat{\Delta}-\Delta|/|\Delta|$), root mean square error (RMSE), and empirical coverage probability for 95\% uniform confidence bands. The results show a general improvement as the sample size $N$ increases. Specifically, the RMSE consistently decreases with $N$ for all estimators. The bias ratios also tend to decrease, although not always monotonically across all quantile points examined. This overall pattern indicates satisfactory finite-sample performance and suggests convergence of the estimators. Furthermore, the empirical coverage rates of the 95\% uniform confidence bands approach the nominal 95\% level as the sample size grows, supporting the validity of the resampling inference procedure.

\begin{table}[h]
	\centering
	\fontsize{10}{15}\selectfont
	\begin{threeparttable}
		\caption{Simulated Statistics of the Local Treatment Effects}\label{tb:simu}
		\begin{tabular}{l@{\hspace{1em}} l@{\hspace{2em}} r@{\hspace{1em}} r@{\hspace{1em}} r@{\hspace{2em}} r@{\hspace{1em}} r r@{\hspace{2em}} r@{\hspace{1em}} r@{\hspace{1em}} r}
			\hline
			& & \multicolumn{3}{l}{Bias Ratio} & \multicolumn{3}{l}{RMSE} & \multicolumn{3}{l}{Unif Cvg} \\
			\cmidrule(r{1.5em}){3-5} \cmidrule(r{1.5em}){6-8} \cmidrule(r){9-11} 
			\(N=\) & & 1000 & 2000 & 4000 & 1000 & 2000 & 4000  & 1000 & 2000 & 4000 \\
			\hline
			\(\Delta_\mu\) & & 0.029 & 0.023 & 0.023  & 0.118 & 0.084 & 0.061 & 0.953 & 0.953 & 0.946 \\
			\hline
			$\Delta_{Id}(y)$ 
			& \(y = y_{0.1}\) &  0.069 & 0.022 & 0.111 &  3.595 & 2.820 & 2.248 & 0.977 & 0.979 & 0.978 \\
			& \(y = y_{0.2}\) &  0.183 & 0.204 & 0.132 &  4.087 & 3.225 & 2.605 & & &   \\
			& \(y = y_{0.3}\) &  0.054 & 0.049 & 0.098 &  4.577 & 3.569 & 2.877 & & &    \\
			& \(y = y_{0.4}\) &  0.027 & 0.014 & 0.049 &  4.947 & 3.800 & 3.126 & & &    \\
			& \(y = y_{0.5}\) &  0.044 & 0.055 & 0.012 &  5.137 & 4.081 & 3.384 & & &    \\
			& \(y = y_{0.6}\) &  0.070 & 0.039 & 0.018 &  5.228 & 4.258 & 3.481 & & &    \\
			& \(y = y_{0.7}\) &  0.057 & 0.049 & 0.008 &  5.026 & 4.129 & 3.388 & & &   \\
			& \(y = y_{0.8}\) &  0.104 & 0.057 & 0.027 &  4.459 & 3.696 & 2.980 & & &    \\
			& \(y = y_{0.9}\) &  0.068 & 0.033 & 0.002 &  2.956 & 2.442 & 2.029 & & &   \\
			\hline
			\(\Delta_{Q}(\tau)\) 
			&\(\tau = 0.10\) &  0.190 & 0.172 & 0.154 &  0.182 & 0.131 & 0.096 &  0.892 & 0.938 & 0.967 \\
			&\(\tau = 0.20\) &  0.073 & 0.066 & 0.069 &  0.169 & 0.123 & 0.089 & &  &    \\
			&\(\tau = 0.30\) &  0.033 & 0.031 & 0.037 &  0.165 & 0.120 & 0.086 & &  &    \\
			&\(\tau = 0.40\) &  0.018 & 0.019 & 0.020 &  0.164 & 0.118 & 0.086 & &  &    \\
			&\(\tau = 0.50\) &  0.007 & 0.010 & 0.011 &  0.162 & 0.117 & 0.086 & &  &    \\
			&\(\tau = 0.60\) &  0.003 & 0.001 & 0.004 &  0.162 & 0.118 & 0.087 & &  &    \\
			&\(\tau = 0.70\) &  0.009 & 0.004 & 0.001 &  0.164 & 0.117 & 0.086 & &  &    \\
			&\(\tau = 0.80\) &  0.009 & 0.009 & 0.002 &  0.170 & 0.122 & 0.086 & &  &    \\
			&\(\tau = 0.90\) &  0.007 & 0.009 & 0.002 &  0.190 & 0.132 & 0.091 & &  &    \\
			\hline
			\(\Delta_L(\tau)\) 
			&\(\tau = 0.10\) &  0.003 & 0.005 & 0.014 &  0.014 & 0.010 & 0.007  & 0.871 & 0.917 & 0.945 \\
			&\(\tau = 0.20\) &  0.006 & 0.003 & 0.003 &  0.023 & 0.016 & 0.011 & &  &    \\
			&\(\tau = 0.30\) &  0.000 & 0.000 & 0.002 &  0.030 & 0.021 & 0.014 & &  &    \\
			&\(\tau = 0.40\) &  0.009 & 0.006 & 0.007 &  0.036 & 0.025 & 0.017 & &  &    \\
			&\(\tau = 0.50\) &  0.022 & 0.014 & 0.015 &  0.041 & 0.029 & 0.019 & &  &    \\
			&\(\tau = 0.60\) &  0.041 & 0.027 & 0.026 &  0.046 & 0.031 & 0.021 & &  &    \\
			&\(\tau = 0.70\) &  0.073 & 0.048 & 0.044 &  0.054 & 0.034 & 0.022 & &  &    \\
			&\(\tau = 0.80\) &  0.130 & 0.086 & 0.074 &  0.056 & 0.036 & 0.023 & &  &    \\
			&\(\tau = 0.90\) &  0.271 & 0.186 & 0.145 &  0.059 & 0.037 & 0.023 & &  &   \\
			\hline
		\end{tabular}
	\end{threeparttable}
\end{table}

\section{Conclusion} \label{sec:conclusion}
This paper develops a unified framework for the identification, estimation, and uniform inference of a class of local treatment effects (LTEs) in the sharp regression kink design. The identification strategy applies to Hadamard differentiable functionals of the outcome distribution and utilizes a continuity assumption on the conditional density of the outcome variable at the kink point. This approach yields a unified identification formula and establishes a connection between the LTE and the local average structural derivative.

For estimation and inference, we consider two classes of local polynomial constrained regression estimators and develop their asymptotic theory, including a resampling method for uniform inference. The framework covers parameters such as effects on the mean, the distribution, and inequality measures. We apply the methods to re-examine the effect of unemployment insurance on unemployment durations, estimating the local distributional and Lorenz treatment effects. This analysis provides information on the impacts of the policy on the shape and dispersion of the duration distribution.
An extension of this framework to the fuzzy regression kink design could be a direction for future research, potentially broadening its applicability.

\renewcommand{\thesection}{A}
\renewcommand{\theequation}{A.\arabic{equation}}
\setcounter{equation}{0}
\begin{appendices}
\section{Regularity Conditions and Auxiliary Lemmas} \label{sec:apdx}
Recall the notation: 
Let \(I_{x_0}^o:=I_{x_0}\backslash\{x_0\}\) be a deleted neighborhood of the kink \(x_0\). Define \(b_0:=b(x_0)\) and \(y_\tau := Q_{Y|X}(\tau|x_0)\).
Define \(h(x,e):=g(b(x),x,e)\), and \(h_x(x,e):=\frac{\partial}{\partial x}h(x,e)=b'(x)g_1(b(x),x,e) + g_2(b(x),x,e)\).
Let $\mathcal{V}(y,x) := \{e \in \mathbb{R}^{d_\varepsilon}: h(x,e) \leq y\}$ be the volume of unobservables yielding an outcome less than or equal to $y$, and  \(\partial \mathcal{V}(y,x):=\{e \in \mathbb{R}^{d_\varepsilon}: h(x,e)= y\}\)  denote its boundary.

\begin{assumptionR}[Regularity Conditions for Identification]~\label{A:id.regular}
	\begin{itemize}
		\item[(i)]
		\(g_1(b_0,x_0,\cdot)\) is measurable and satisfies 
		\[
		P\left[\big| g(b_0+\delta,x_0,\varepsilon) - g(b_0,x_0,\varepsilon) - \delta g_1(b_0,x_0,\varepsilon) \big| \geq c |\delta| \,\middle|\, X=x_0\right]  = o(\delta)
		\]
		as \(\delta \to 0\) for all \(c > 0\).

		\item[(ii)]
		The conditional distribution of \((Y,g_1(b_0,x_0,\varepsilon))\) given \(X=x_0\) is absolutely continuous with respect to Lebesgue measure, and its density \(f_{Y,g_1|X}(y,y'|x_0)\) is continuous in \(y\) for all \(y'\). Moreover, there exists a Lebesgue integrable function \(\varpi:\mathbb{R} \to \mathbb{R}\) with \(\int |y'\varpi(y')|\,dy' < \infty\) such that for all \((y,y')\), \(f_{Y,g_1|X}(y,y'|x_0) \leq |\varpi(y')|\).

		\item[(iii)]
		The conditional distribution $x\mapsto F_{Y|X}(y|x)$ is continuously differentiable on $I^o_{x_0}$ for all \(y \in \mathcal{Y}_x\). The left and right limits of this derivative at $x_0$, \(y\mapsto \partial F_{Y|X}(y|x_0^-)/\partial x\) and \(y\mapsto \partial F_{Y|X}(y|x_0^+)/\partial x\), are continuous on \(\mathcal{Y}_{x_0}\). Moreover, for all $x\in I_{x_0}^o$ and $t>0$ such that $x+t \in I_{x_0}^o$, $h(x,\cdot)$ is measurable and satisfies
		\begin{align*}
			P\left[\big| h(x+t, \varepsilon) - h(x, \varepsilon)-
			t \, h_x(x,\varepsilon) \big| \geq c \, t  \,\middle|\, X=x\right] = o(t)
		\end{align*}
		as $t \to 0^+$ for all $c > 0$.
		
		\item[(iv)]
		The conditional distribution of $(Y, h_x(x,\varepsilon))$ given $X=x$ is absolutely continuous with respect to the Lebesgue measure for each $x \in I_{x_0}^o$. Its conditional density, $f_{Y,h_x|X}(y,y'|x)$, satisfies the following conditions:
		(a) The function $(y, y') \mapsto f_{Y,h_x|X}(y,y'|x)$ is continuous for all $x \in I_{x_0}^o$.
		(b) There exists a Lebesgue integrable function $\varphi: \mathbb{R} \to \mathbb{R}$ such that $\int |y'\varphi(y')|\,dy' < \infty$ and \(f_{Y,h_x|X}(y,y'|x) \leq |\varphi(y')|\)  for all \((y,y')\) and \(x\in I_{x_0}^o\).
		
		\item[(v)]
		The conditional distribution of $\varepsilon$ given $(Y,X)$ is absolutely continuous with respect to the Lebesgue measure. Its conditional density, $f_{\varepsilon|Y,X}(e|y,x)$, satisfies the following conditions:
		(a) The function $(y,x) \mapsto f_{\varepsilon|Y,X}(e|y,x)$ is continuous on $\mathcal{Y}_{x} \times I_{x_0}$ for all $e\in \mathcal{E}$.
		(b) There exists a Lebesgue integrable function $\chi:\mathcal{E} \to \mathbb{R}$ and positive constants $C_1, C_2$ such that for all $(e,y,x)$:
		$f_{\varepsilon|Y,X}(e|y,x) \leq C_1|\chi(e)|$ for $(y,x) \in \mathcal{Y}_{x} \times I_{x_0}$, and
		$|\partial f_{\varepsilon|X}(e|x)/\partial x| \leq C_2|\chi(e)|$ for $x \in I_{x_0}$.
	\end{itemize}
\end{assumptionR}
Assumption \ref{A:id.regular} collects several technical regularity conditions. Parts (i) and (iii), analogous to Assumption A3 in \cite{hoderlein2007identification}, impose smoothness on the structural marginal effects, $g_1$ and $g_2$. Parts (ii), (iv), and (v) are standard moment and boundedness conditions required to justify the application of the Lebesgue dominated convergence theorem in our proofs.

We now restate a result from \cite{chiang2019causal} concerning the asymptotic properties of the QRKD estimator. This result provides a basis for the asymptotic theory of our composite RKD estimator.
\begin{lemmaA}[\citealp{chiang2019causal}, Corollary 2]\label{lem:asy.qrkd.cs19}
	Suppose Assumptions \ref{A:sharp.kink} and \ref{A:asy.cs19} hold. Then,
		\begin{align*}
			\sqrt{nh_{n,\tau}^3}\left(\widehat{\qrkd}(\tau) - \qrkd(\tau)\right)
			\leadsto 
			\G^Q(\tau) := \frac{\mathbb{Z}^Q(\tau,2)- \mathbb{Z}^Q(\tau,3)}{b'(x_0^+)-b'(x_0^-)},
		\end{align*}
		uniformly in \(\tau \in \mathcal{T}\subset (0,1)\), where  \(\mathbb{Z}^Q(\cdot,2)\) and \(\mathbb{Z}^Q(\cdot,3)\) are tight zero-mean Gaussian processes of the normalized quantile derivative processes.
		\(\G^Q:\varOmega \to \ell^\infty(\mathcal{T})\) is a tight zero-mean Gaussian process with covariance function 
		\begin{align*}
			E\left[\G^Q(\tau_1)\G^Q(\tau_2)\right] 
			=  
			\frac{(\tau_1 \wedge \tau_2 - \tau_1 \tau_2)\cdot  (\iota_{2}-\iota_{3})^\top  \widebar{\varGamma}_p^{-1} T(\tau_1,\tau_2) \widebar{\varGamma}_p^{-1} (\iota_{2}-\iota_{3}) }{(b'(x_0^+)-b'(x_0^-))^2 f_X(x_0) f_{Y|X}(y_{\tau_1}|x_0)f_{Y|X}(y_{\tau_2}|x_0)}
		\end{align*}
		for all \(\tau_1, \tau_2 \in \mathcal{T}\), where \(y_\tau:=Q_{Y|X}(\tau|x_0)\)
		, and  
		\[T(\tau_1,\tau_2):=\frac{1}{\sqrt{c(\tau_1)c(\tau_2)}}\int_{\mathbb{R}} \bar{r}_p\left(\frac{u}{c(\tau_1)}\right) \bar{r}_p\left(\frac{u}{c(\tau_2)}\right)^\top  K\left(\frac{u}{c(\tau_1)}\right) K\left(\frac{u}{c(\tau_2)}\right)\,du.\]
\end{lemmaA}

\begin{lemmaA}[First Stage Consistency]\label{lem:asy.first.stage}
	Let \[\hat{\varepsilon}^m(y,x,\theta):=\left(\varphi(y,\theta) - \bar{r}_p(x-x_0)\hat{\alpha}(\theta)\right)\I\left(\left|x-x_0\right|\leq h_{n,\theta}\right).\]
	Under Assumption \ref{A:asy}, \(\hat{\varepsilon}^m(y,x,\theta) = \varepsilon^m(y,x,\theta)\I\left(\left|x-x_0\right|\leq h_{n,\theta}\right) + o_{P}(1)\) uniformly in \(\theta \in \widebar{\varTheta}\).
\end{lemmaA}

\section{Bandwidth Selection} \label{sec:apdx.bandwidth}
We first recall the necessary notation. Let $\bar{r}_p(u):=(1, u \delta_u^+, u \delta_u^-, \dots, u^p \delta_u^+, u^p \delta_u^-)^\top$ be the $(2p+1)$-dimensional vector of regressors, where $\delta_u^+:= \I(u\geq 0)$ and $\delta_u^-:= \I(u<0)$. The following kernel-dependent constants are used in the derivation:  $\widebar{\varGamma}_p:= \int_{\mathbb{R}} \bar{r}_p(u)\bar{r}_p(u)^\top K(u)\,du$; $\bar{\vartheta}_{p,q}^\pm := \int_{\mathbb{R}_\pm}\bar{r}_p(u) u^{q} K(u)\,du$; $\widebar{\varPsi}_p^\pm:=\int_{\mathbb{-}_\pm} \bar{r}_p(u)\bar{r}_p(u)^\top K^2(u)\,du$; $\widebar{\varPsi}_p:=\int_{\mathbb{R}} \bar{r}_p(u)\bar{r}_p(u)^\top K^2(u)\,du$.

\subsection{Bandwidth Selector for Estimating $\rkd_m$}
This section details the procedure for selecting a pointwise mean squared error (MSE)-optimal bandwidth for estimating the difference in derivatives, $m^{(1)}(\theta,x_0^+) - m^{(1)}(\theta,x_0^-)$, using our constrained regression estimator.
Under Assumption \ref{A:asy}(iv), the asymptotic MSE (AMSE) of \(\hat{m}^{(1)}(\theta,x_0^+) - \hat{m}^{(1)}(\theta,x_0^-)\) is given by
\begin{align*}
	AMSE\left[\hat{m}^{(1)}(\theta,x_0^+) - \hat{m}^{(1)}(\theta,x_0^-)\right]
	&=\left(ABias\left[\hat{m}^{(1)}(\theta,x_0^+) - \hat{m}^{(1)}(\theta,x_0^-)\right]\right)^2\\
	&\quad + AVar\left[\hat{m}^{(1)}(\theta,x_0^+) - \hat{m}^{(1)}(\theta,x_0^-)\right]\\
	&=h_{n,\theta}^{2p}\mathsf{B}^2_{1,p}(\theta) + \frac{1}{nh_{n,\theta}^3}\mathsf{V}_{1,p}(\theta),
\end{align*} 
where  for every  \(\nu, p\in \mathbb{N}_+\) with \(1 \leq \nu \leq p\), 
\begin{align}
	\mathsf{B}_{\nu,p}(\theta)
	&:= \frac{\iota_{2v}^\top \widebar{\varGamma}_p^{-1} \left[m^{(p+1)}(\theta,x_0^+) \bar{\vartheta}_{p,p+1}^+ + m^{(p+1)}(\theta,x_0^-)\bar{\vartheta}_{p,p+1}^-\right]}{(p+1)!} \notag\\ 
	&\qquad - \frac{\iota_{2v+1}^\top \widebar{\varGamma}_p^{-1} \left[m^{(p+1)}(\theta,x_0^+) \bar{\vartheta}_{p,p+1}^+ + m^{(p+1)}(\theta,x_0^-)\bar{\vartheta}_{p,p+1}^-\right]}{(p+1)!} \label{eq:bw.m.B} \\
	\mathsf{V}_{\nu,p}(\theta)
	&:= \frac{\iota_{2v}^\top \widebar{\varGamma}_p^{-1} \left[\sigma_{\varepsilon^m}(\theta,\theta|x_0^+)\widebar{\varPsi}_{p}^+ + \sigma_{\varepsilon^m}(\theta,\theta|x_0^-)\widebar{\varPsi}_{p}^-\right] \widebar{\varGamma}_p^{-1} \iota_{2v}}{f_X(x_0)} \notag\\
	&\qquad + \frac{\iota_{2v+1}^\top \widebar{\varGamma}_p^{-1} \left[\sigma_{\varepsilon^m}(\theta,\theta|x_0^+)\widebar{\varPsi}_{p}^+ + \sigma_{\varepsilon^m}(\theta,\theta|x_0^-)\widebar{\varPsi}_{p}^-\right] \widebar{\varGamma}_p^{-1} \iota_{2v+1}}{f_X(x_0)}  \label{eq:bw.m.V}\\
	&\qquad - \frac{2 \iota_{2v}^\top \widebar{\varGamma}_p^{-1} \left[\sigma_{\varepsilon^m}(\theta,\theta|x_0^+)\widebar{\varPsi}_{p}^+ + \sigma_{\varepsilon^m}(\theta,\theta|x_0^-)\widebar{\varPsi}_{p}^-\right] \widebar{\varGamma}_p^{-1} \iota_{2v+1}}{f_X(x_0)}. \notag	
\end{align}
 
 Following the general approach for MSE-optimal bandwidth selection in local polynomial regression (e.g., \cite{calonico2014robust}), we can derive the optimal bandwidth for our constrained estimator. For a general problem of estimating a $\nu$th-order derivative ratio with a $p$th-order local polynomial, the MSE-optimal bandwidth takes the form:
\begin{align*}
	h_{\nu,p}(\theta) 
	&= \arg\min_{h>0}\left\{h^{2(p+1-\nu)}\mathsf{B}^2_{\nu,p}(\theta)+\frac{1}{nh^{1+2\nu}}\mathsf{V}_{\nu,p}(\theta)\right\}\\
	&= \left\{\frac{1+2\nu}{2(p+1-\nu)}\,\frac{\mathsf{V}_{\nu,p}(\theta)}{\mathsf{B}^2_{\nu,p}(\theta)}\right\}^{1/(2p+3)}  n^{-1/(2p+3)}.
\end{align*}
Here, $\mathsf{V}_{\nu,p}(\theta)$ and $\mathsf{B}_{\nu,p}(\theta)$ represent the leading terms of the asymptotic variance and bias, respectively.
Our specific estimand, $\rkd_m(\theta)$, is a ratio of first-order derivatives, which corresponds to the case $\nu=1$. Therefore, the pointwise MSE-optimal bandwidth for our $p$th-order estimator is given by $h_{n,\theta}=h_{1,p}(\theta)$. Implementing this choice in practice requires a feasible procedure to estimate the unknown variance and bias terms. The following algorithm details these steps.
\begin{myalgorithm}[Plug-in Bandwidth Selector for Estimating \(\rkd_m\)]~\label{alg:bandwidth.m}\\
	Fix positive integers $p,q$ such that $p < q$.
	\begin{enumerate}[\itshape Step 1.]

		\item (Pilot Bandwidths)
		The pilot bandwidth \(\hat{b}_{n,\theta}\) is given by
		\begin{align*}
			\hat{b}_{n,\theta} 
			= \left\{ \frac{2p+3}{2(q-p)}\frac{\widetilde{\mathsf{V}}_{p+1,q}(\theta)}{\widetilde{\mathsf{B}}_{p+1,q}^2(\theta)}\right\}^{1/(2q+3)}  n^{-1/(2q+3)}, \quad \theta \in \widebar{\varTheta}.
		\end{align*}
		Calculating this requires an estimate of $\widetilde{\mathsf{B}}_{p+1,q}$ and $\widetilde{\mathsf{V}}_{p+1,q}$, which in turn depends on the $(q+1)$-th order derivatives, $\tilde{m}^{(q+1)}(\theta,x_0^\pm)$, and the conditional variance of the regression error, $\sigma_{\varepsilon^m}(\theta,\theta|x_0^\pm)$, respectively. To obtain these high-order derivative estimates, we fit a $(q+1)$-th order global polynomial regression using our constrained specification:
		\begin{align*}
			\tilde{\gamma}_{q+1}(\theta) = \arg\min_{\gamma\in\mathbb{R}^{2q+3}}\sum_{i=1}^{n}\left(\varphi(Y_i,\theta)-\bar{r}_{q+1}(X_i-x_0)^\top\gamma\right)^2.
		\end{align*}
		The required derivative estimates are then extracted from the resulting coefficient vector:
		$\tilde{m}^{(q+1)}(\theta,x_0^+)=\iota_{2q+2}^\top\tilde{\gamma}_{q+1}(\theta)/(q+1)!$ and $\tilde{m}^{(p+1)}(\theta,x_0^-)=\iota_{2q+3}^\top\tilde{\gamma}_{q+1}(\theta)/(q+1)!$.
		
		Next, we estimate this conditional variance $\sigma_{\varepsilon^m}(\theta,\theta|x_0^\pm)$ using a two-stage procedure. First, we obtain the fitted residuals from the global polynomial regression in the previous step: \(\tilde{\varepsilon}^m(y,x,\theta)=\varphi(y,\theta)-\bar{r}_{q+1}(x-x_0)^\top\tilde{\gamma}_{q+1}(\theta)\). Second, we estimate the conditional mean at $x_0$ from the left and right by running separate local linear regressions:
		\begin{align*}
			\tilde{\sigma}_{\varepsilon^m}(\theta,\theta|x_0^\pm)=\iota_1^\top\arg\min_{\kappa\in\mathbb{R}^3}\sum_{i=1}^{n}\delta_i^\pm\left([\tilde{\varepsilon}^m(Y_i,X_i,\theta)]^2 - (1,X_i-x_0)^\top \kappa\right)^2.
		\end{align*}
		The density $f_X(x_0)$ is estimated by 
		$$\hat{f}_X(x_0) = \frac{1}{nv_n}\sum_{i=1}^{n} K\left(\frac{X_i-x_0}{v_n}\right),$$ 
		where $v_n= 2.576 \cdot\min\left\{sd(X), IQR(X)/1.349 \right\} \cdot n^{-1/5}$; $sd$ denotes the sample standard error and $IQR$ denotes the interquartile range.  

		\item (Main Bandwidths)
		The main bandwidth \(\hat{h}_{n,\theta}\) is given by
		\begin{align*}
			\hat{h}_{n,\theta} = \hat{h}_{1,p}(\theta) 
			=\left\{\frac{3}{2p}\,\frac{\widehat{\mathsf{V}}_{1,p}(\theta)}{\widehat{\mathsf{B}}_{1,p}^2(\theta)}\right\}^{1/(2p+3)}  n^{-1/(2p+3)}, \quad \theta \in \widebar{\varTheta}.
		\end{align*}
		To estimate the $(p+1)$th-order derivatives needed for the bias term, we fit a $(p+1)$th-order local polynomial constrained regression, using the pilot bandwidth $\hat{b}_{n,\theta}$:
		\begin{align*}
			\hat{\gamma}_{p+1}(\theta) = \arg\min_{\gamma\in\mathbb{R}^{2p+3}}\sum_{i=1}^{n}\left(\varphi(Y_i,\theta)-\bar{r}_{p+1}(X_i-x_0)^\top\gamma\right)^2K\left(\frac{X_i-x_0}{\hat{b}_{n,\theta}}\right).
		\end{align*}
		The required derivatives, $\hat{m}^{(p+1)}(\theta,x_0^\pm)$, are then extracted from the coefficient vector: \(\hat{m}^{(p+1)}(\theta,x_0^+)=\iota_{2p+2}^\top\hat{\gamma}_{p+1}(\theta)/(p+1)!\), and \(\hat{m}^{(p+1)}(\theta,x_0^-)=\iota_{2p+3}^\top\hat{\gamma}_{p+1}(\theta)/(p+1)!\).
		
		To estimate the term  \(\sigma_{\varepsilon^m}(\theta,\theta|x_0^\pm)\) in \(\widehat{\mathsf{V}}_{1,p}\), we first obtain fitted residuals from the regression in the step above: \(\hat{\varepsilon}^m(y,x,\theta)=\varphi(y,\theta)-\bar{r}_{p+1}(x-x_0)^\top\hat{\gamma}_{p+1}(\theta)\). We then estimate the conditional error variance, $\hat{\sigma}_{\varepsilon^m}(\theta,\theta|x_0^\pm)$, by running separate local linear regressions of the squared residuals on each side of the kink, again using the pilot bandwidth $\hat{b}_{n,\theta}$:
		\begin{align*}
			\hat{\sigma}_{\varepsilon^m}(\theta,\theta|x_0^\pm)=\iota_1^\top\arg\min_{\kappa\in\mathbb{R}^3}\sum_{i=1}^{n}\delta_i^\pm\left([\hat{\varepsilon}^m(Y_i,X_i,\theta)]^2 - (1,X_i-x_0)^\top \kappa\right)^2K\left(\frac{X_i-x_0}{\hat{b}_{n,\theta}}\right).
		\end{align*}
		The estimator of the density $\hat{f}_X(x_0)$ is the same as that used in Step 1.
	\end{enumerate}
\end{myalgorithm}

\subsection{Bandwidth Selector for Estimating the LLTE $\Delta_L$}
Deriving the fully optimal bandwidth for the composite local Lorenz treatment effect estimator is complex. We therefore adopt a practical approximation based on the standard MSE-optimal formula for a first-order derivative:
\begin{align*}
	h^L_{n,\tau} = \left\{\frac{3}{2p}\frac{\mathsf{V}^L_{1,p}(\tau)}{[ \mathsf{B}^L_{1,p}(\tau)]^2}\right\}^{1/(2p+3)} n^{-1/(2p+3)}, \quad \tau \in \mathcal{T},
\end{align*}
where, for \(\nu, p\in \mathbb{N}_+\) with \(1\leq \nu \leq p\),
\begin{align*}
	\mathsf{B}^L_{\nu,p}(\tau)
	&:=\frac{1}{\mu_0}\left(\int_{0}^{\tau}\mathsf{B}^Q_{\nu,p}(u)\,du - L_{Y|X}(\tau|x_0)\mathsf{B}^\mu_{\nu,p}\right),\\
	\mathsf{V}^L_{\nu,p}(\tau)
	&:=\frac{1}{\mu^2_0}\left(\int_{0}^{\tau}\mathsf{V}^Q_{\nu,p}(u)\,du + L_{Y|X}^2(\tau|x_0) \mathsf{V}^\mu_{\nu,p}\right).
\end{align*}
The terms $\mathsf{B}^L_{1,p}$ and $\mathsf{V}^L_{1,p}$ are constructed from the asymptotic bias and variance components of the underlying mean and quantile estimators. Specifically, the components related to the mean effect, $\mathsf{B}^\mu_{1,p}$ and $\mathsf{V}^\mu_{1,p}$, are as defined in Equations (\ref{eq:bw.m.B}) and (\ref{eq:bw.m.V}) for the case where $m(x,\theta) = E[Y|X=x]$.  The components related to the quantile effect, $\mathsf{B}^Q_{1,p}$ and $\mathsf{V}^Q_{1,p}$, represent the leading bias and variance terms, respectively, of the estimated difference in quantile derivatives, $\widehat{Q}^{(1)}_{Y|X}(\tau|x_0^+)-\widehat{Q}^{(1)}_{Y|X}(\tau|x_0^-)$. They are given by:
\begin{align*}
	\mathsf{B}^Q_{\nu,p}(\tau)
	&=\frac{\iota_{2v}^\top \widebar{\varGamma}_p^{-1} \left(Q^{(p+1)}_{Y|X}(\tau|x_0^+) \bar{\vartheta}_{p,p+1}^+ + Q^{(p+1)}_{Y|X}(\tau|x_0^-)\bar{\vartheta}_{p,p+1}^-\right)}{(p+1)!} \notag\\ 
	&\qquad - \frac{\iota_{2v+1}^\top \widebar{\varGamma}_p^{-1} \left(Q^{(p+1)}_{Y|X}(\tau|x_0^+) \bar{\vartheta}_{p,p+1}^+ + Q^{(p+1)}_{Y|X}(\tau|x_0^-)\bar{\vartheta}_{p,p+1}^-\right)}{(p+1)!},\\
	\mathsf{V}^Q_{\nu,p}(\tau)
	&=\frac{\tau(1-\tau)\left[\iota_{2v}^\top \widebar{\varGamma}_p^{-1}\widebar{\varPsi}_p\widebar{\varGamma}_p^{-1}\iota_{2v} + \iota_{2v+1}^\top \widebar{\varGamma}_p^{-1}\widebar{\varPsi}_p\widebar{\varGamma}_p^{-1}\iota_{2v+1} - 2\iota_{2v}^\top \widebar{\varGamma}_p^{-1}\widebar{\varPsi}_p\widebar{\varGamma}_p^{-1}\iota_{2v+1}\right]}{f_X(x_0)f_{Y|X}(y_\tau|x_0)}.
\end{align*}

The following algorithm details the bandwidth selection procedure for the local Lorenz treatment effect.
\begin{myalgorithm}[Plug-in Bandwidth Selector for Estimating \(\Delta_L\)]\label{alg:bandwidth.lorenz} ~\\
	Fix positive integers $p,q$ such that $p < q$. Define a fine grid of quantile levels, e.g., $\mathcal{U} = \{0.01, 0.02, \dots, 0.99\}$.
	\begin{enumerate}[\itshape Step 1.]
		\item (Pilot Bandwidths)
		The pilot bandwidths are given by
		\begin{align*}
			\hat{b}_n 
			&= \left\{ \frac{2p+3}{2(q-p)} \frac{\widehat{\mathsf{V}}^\mu_{p+1,q}}{[\widehat{\mathsf{B}}^\mu_{p+1,q}]^2}\right\}^{1/(2q+3)}n^{-1/(2q+3)},
			\\
			\text{and }\; \hat{b}_{n,u} 
			&= \left\{ \frac{2p+3}{2(q-p)} \frac{\widehat{\mathsf{V}}^Q_{p+1,q}(u)}{[\widehat{\mathsf{B}}^Q_{p+1,q}(u)]^2}\right\}^{1/(2q+3)} n^{-1/(2q+3)}, \quad u \in \mathcal{U}.
		\end{align*} 
		The terms $\widehat{\mathsf{B}}^\mu_{p+1,q}$ and $\widehat{\mathsf{V}}^\mu_{p+1,q}$ are independent of $u$ and are computed once following the procedure for the conditional mean in Algorithm \ref{alg:bandwidth.m}.

		To estimate the $(q+1)$th-order derivatives $\widehat{Q}^{(q+1)}_{Y|X}(u|x_0^\pm)$ needed for the quantile bias term $\widehat{\mathsf{B}}^Q_{p+1,q}$. We fit a $(q+1)$th-order global constrained polynomial quantile regression to obtain the coefficient vector:
		\begin{align*}
			\hat{\gamma}_{q+1}(u) = \arg\min_{\gamma\in\mathbb{R}^{2q+3}}\sum_{i=1}^{n}\rho_u\left(Y_i-\bar{r}_{q+1}(X_i-x_0)^\top\gamma\right).
		\end{align*}
		Then,  we extract the derivative estimates: $\hat{y}_u = \iota_1^\top \hat{\gamma}_{q+1}(u)$, \(\widehat{Q}^{(q+1)}_{Y|X}(u|x_0^+)=\iota_{2q+2}^\top\hat{\gamma}_{q+1}(u)/(q+1)!\), and \(\widehat{Q}^{(q+1)}_{Y|X}(u|x_0^-)=\iota_{2q+3}^\top\hat{\gamma}_{q+1}(u)/(q+1)!\).
		
		Next, the (conditional) densities \(f_X\) and \(f_{Y|X}\) in \(\widehat{\mathsf{V}}^Q_{p+1,q}(u)\) are estimated as follows:
		\begin{align}
			\begin{split}
				\hat{f}_X(x_0) 
				&= \frac{1}{nv_n}\sum_{i=1}^{n} K\left(\frac{X_i-x_0}{v_n}\right), \\
				\text{and }\; 
				\hat{f}_{Y|X}(y|x_0) 
				&= \frac{1}{h_{1,n}} \frac{\sum_{i=1}^{n}K\left(\frac{Y_i-y}{h_{1,n}}\right)K\left(\frac{X_i-x_0}{h_{2,n}}\right)}{\sum_{i=1}^{n}K\left(\frac{X_i-x_0}{h_{2,n}}\right)},
			\end{split}  \label{eq:esti.densities}
		\end{align}
		where $v_n$ is given by Step 1 of Algorithm \ref{alg:bandwidth.m}. We select the bandwidths $h_{1,n}$ and $h_{2,n}$ in $\hat{f}_{Y|X}$ following the rule-of-thumb procedure proposed by \citet[Section 2.2]{bashtannyk2001bandwidth}:
		\begin{align*}
			h_{2,n}
			&=\left\{%
			\frac{16c R^2_K \sigma^5_Y[288 \pi^9 \sigma_X^{58} \lambda^2(c)]^{1/8}}%
			{\varrho_K^4 b^{5/2}v^{3/4}(c)\{v^{1/2}(c)+b[18\pi\sigma_X^{10}\lambda^2(c)]^{1/4}\}}
			\right\}^{1/6} n^{-1/6},\\
			\text{and }\; h_{1,n} &= \left\{%
			\frac{b^2v(c)}{3\sqrt{2\pi}\sigma_X^5\lambda(c)}
			\right\}^{1/4}h_{2,n},
		\end{align*}
		where $\sigma_Y$ and  $\sigma_X$ denote the sample standard deviation; $\lambda(c)=\int_{-c}^{c} \varphi(t)\,dt$; $v(c)=\sqrt{2\pi}\sigma_X^3 (3q \sigma_X^2 + 8 \sigma_Y^2)\lambda(c)-16c\sigma_X^2 \sigma_Y^2 e^{-k^2/2}$; $q$ is the slope of an OLS regression of $Y_i$ on $[1, X_i]$. $R_K=\int K^2(u)\,du$ and $\varrho_K = \int u^2K(u) \,du$. Typically, constant $c$ can be set to 2 or 3.
		
		\item 
		(Main Bandwidths)
		The main bandwidth is given by
		\begin{align*}
			\hat{h}^L_{n,\tau} = \left\{\frac{3}{2p}\frac{\widehat{\mathsf{V}}^L_{1,p}(\tau)}{[ \widehat{\mathsf{B}}^L_{1,p}(\tau)]^2}\right\}^{1/(2p+3)} n^{-1/(2p+3)}, \quad \tau\in\mathcal{T}
		\end{align*}
		where \(\widehat{L}_{Y|X}(\tau|x_0) = \int_{0}^{\tau} \hat{y}_u \,du/\widehat{\mu}_0\);
		\begin{align*}
			\widehat{\mathsf{B}}^L_{1,p}(\tau)
			&=\frac{1}{\mu_0}\left(\int_{0}^{\tau}\widehat{\mathsf{B}}^Q_{1,p}(u)\,du - \widehat{L}_{Y|X}(\tau|x_0)\widehat{\mathsf{B}}^\mu_{1,p}\right),\\
			\text{and }\; \widehat{\mathsf{V}}^L_{1,p}(\tau)
			&=\frac{1}{\mu^2_0}\left(\int_{0}^{\tau}\widehat{\mathsf{V}}^Q_{1,p}(u)\,du + \widehat{L}_{Y|X}^2(\tau|x_0) \widehat{\mathsf{V}}^\mu_{1,p}\right).
		\end{align*}
		 The terms related to the mean effect—$\hat{\mu}_0$, $\widehat{\mathsf{B}}^\mu_{1,p}$, and $\widehat{\mathsf{V}}^\mu_{1,p}$—are independent of the quantile level $u$. They are computed once as described in Algorithm \ref{alg:bandwidth.m}, Step 2, using the pilot bandwidth \(\hat{b}_n\).

		The following components are computed for each $u \in \mathcal{U}$:
		To estimate the $(p+1)$-th order derivatives, we fit a $(p+1)$-th order local constrained polynomial quantile regression, using the corresponding pilot bandwidth $\hat{b}_{n,u}$:
		\begin{align*}
			\hat{\gamma}_{p+1}(u) = \arg\min_{\gamma\in\mathbb{R}^{2p+3}}\sum_{i=1}^{n}\rho_\tau\left(Y_i-\bar{r}_{p+1}(X_i-x_0)^\top\gamma\right)K\left(\frac{X_i-x_0}{\hat{b}_{n,u}}\right).
		\end{align*}
		The interpret and derivative estimates are then extracted from the coefficient vector: $\hat{y}_u = \iota_1^\top \hat{\gamma}_{p+1}(u)$, \(\widehat{Q}^{(p+1)}_{Y|X}(u|x_0^+)=\iota_{2p+2}^\top\hat{\gamma}_{p+1}(u)/(p+1)!\), and \(\widehat{Q}^{(p+1)}_{Y|X}(u|x_0^-)=\iota_{2p+3}^\top\hat{\gamma}_{p+1}(u)/(p+1)!\).
		The conditional densities, $f_X$ and $f_{Y|X}$, required for the term \(\widehat{\mathsf{V}}^Q_{1,p}(u)\) are estimated using standard kernel methods as detailed in Equation (\ref{eq:esti.densities}).
	\end{enumerate}
\end{myalgorithm}

\begin{remark}
	The procedure in Algorithm \ref{alg:bandwidth.lorenz} can be straightforwardly adapted to find the MSE-optimal bandwidth for the standard QRKD estimator, $\widehat{\qrkd}(\tau)$. The steps are analogous. For each quantile of interest $\tau \in \mathcal{T}$: (1) A pilot bandwidth, $\hat{b}_{n,\tau}$, is computed as described in Step 1 of the algorithm. (2) The bias and variance components, $\widehat{\mathsf{B}}^Q_{1,p}(\tau)$ and $\widehat{\mathsf{V}}^Q_{1,p}(\tau)$, are computed as described in Step 2, using the pilot bandwidth $\hat{b}_{n,\tau}$ from the previous step. These components are then plugged into the standard formula for the main optimal bandwidth:
	\begin{align*}
		\hat{h}_{n,\tau} = \left\{\frac{3}{2p}\frac{\widehat{\mathsf{V}}^Q_{1,p}(\tau)}{[ \widehat{\mathsf{B}}^Q_{1,p}(\tau)]^2}\right\}^{1/(2p+3)} n^{-1/(2p+3)}, \quad \tau \in \mathcal{T},
	\end{align*}
	This provides a MSE-optimal bandwidth for the $p$th-order local polynomial constrained quantile regression estimator in (\ref{eq:esti.constrained.qr.reg}). Our result generalizes the bandwidth selector in Appendix C.1 of \citet{chiang2019causal}, which focuses on the local linear ($p=1$) case.
\end{remark}

\end{appendices}

\onehalfspacing
\bibliographystyle{apalike}
\bibliography{teSRKD_ref}

\begin{thebibliography}{}

\bibitem[Bashtannyk and Hyndman, 2001]{bashtannyk2001bandwidth}
Bashtannyk, D.~M. and Hyndman, R.~J. (2001).
\newblock Bandwidth selection for kernel conditional density estimation.
\newblock {\em Computational Statistics \& Data Analysis}, 36(3):279--298.

\bibitem[Bhattacharya, 2007]{bhattacharya2007inference}
Bhattacharya, D. (2007).
\newblock Inference on inequality from household survey data.
\newblock {\em Journal of Econometrics}, 137(2):674--707.

\bibitem[Calonico et~al., 2014]{calonico2014robust}
Calonico, S., Cattaneo, M.~D., and Titiunik, R. (2014).
\newblock Robust nonparametric confidence intervals for
  regression-discontinuity designs.
\newblock {\em Econometrica}, 82(6):2295--2326.

\bibitem[Card et~al., 2015a]{card2015effect}
Card, D., Johnston, A., Leung, P., Mas, A., and Pei, Z. (2015a).
\newblock The effect of unemployment benefits on the duration of unemployment
  insurance receipt: New evidence from a regression kink design in missouri,
  2003--2013.
\newblock {\em American Economic Review}, 105(5):126--130.

\bibitem[Card et~al., 2015b]{card2015inference}
Card, D., Lee, D.~S., Pei, Z., and Weber, A. (2015b).
\newblock Inference on causal effects in a generalized regression kink design.
\newblock {\em Econometrica}, 83(6):2453--2483.

\bibitem[Chen et~al., 2020]{chen2020quantile}
Chen, H., Chiang, H.~D., and Sasaki, Y. (2020).
\newblock Quantile treatment effects in regression kink designs.
\newblock {\em Econometric Theory}, 36(6):1167--1191.

\bibitem[Chen and Fan, 2019]{chen2019identification}
Chen, H. and Fan, Y. (2019).
\newblock Identification and wavelet estimation of weighted ate under
  discontinuous and kink incentive assignment mechanisms.
\newblock {\em Journal of Econometrics}, 212(2):476--502.

\bibitem[Chernozhukov et~al., 2010]{chernozhukov2010quantile}
Chernozhukov, V., Fern{\'a}ndez-Val, I., and Galichon, A. (2010).
\newblock Quantile and probability curves without crossing.
\newblock {\em Econometrica}, 78(3):1093--1125.

\bibitem[Chernozhukov et~al., 2015]{chernozhukov2015nonparametric}
Chernozhukov, V., Fernandez-Val, I., Hoderlein, S., Holzmann, H., and Newey, W.
  (2015).
\newblock Nonparametric identification in panels using quantiles.
\newblock {\em Journal of Econometrics}, 188(2):378--392.

\bibitem[Chiang et~al., 2019]{chiang2019robust}
Chiang, H.~D., Hsu, Y.-C., and Sasaki, Y. (2019).
\newblock Robust uniform inference for quantile treatment effects in regression
  discontinuity designs.
\newblock {\em Journal of Econometrics}, 211(2):589--618.

\bibitem[Chiang and Sasaki, 2019]{chiang2019causal}
Chiang, H.~D. and Sasaki, Y. (2019).
\newblock Causal inference by quantile regression kink designs.
\newblock {\em Journal of Econometrics}, 210(2):405--433.

\bibitem[Dong, 2018]{dong2018jump}
Dong, Y. (2018).
\newblock Jump or kink? regression probability jump and kink design for
  treatment effect evaluation.
\newblock {\em Unpublished manuscript}.

\bibitem[Florens et~al., 2008]{florens2008identification}
Florens, J.-P., Heckman, J.~J., Meghir, C., and Vytlacil, E. (2008).
\newblock Identification of treatment effects using control functions in models
  with continuous, endogenous treatment and heterogeneous effects.
\newblock {\em Econometrica}, 76(5):1191--1206.

\bibitem[Hoderlein et~al., 2017]{hoderlein2017corrigendum}
Hoderlein, S., Holzmann, H., Kasy, M., and Meister, A. (2017).
\newblock Corrigendum: Instrumental variables with unrestricted heterogeneity
  and continuous treatment.
\newblock {\em The Review of Economic Studies}, 84(2):964--968.

\bibitem[Hoderlein and Mammen, 2007]{hoderlein2007identification}
Hoderlein, S. and Mammen, E. (2007).
\newblock Identification of marginal effects in nonseparable models without
  monotonicity.
\newblock {\em Econometrica}, 75(5):1513--1518.

\bibitem[Hoderlein and Mammen, 2009]{hoderlein2009identification}
Hoderlein, S. and Mammen, E. (2009).
\newblock Identification and estimation of local average derivatives in
  non-separable models without monotonicity.
\newblock {\em The Econometrics Journal}, 12(1):1--25.

\bibitem[Imbens and Newey, 2009]{imbens2009identification}
Imbens, G.~W. and Newey, W.~K. (2009).
\newblock Identification and estimation of triangular simultaneous equations
  models without additivity.
\newblock {\em Econometrica}, 77(5):1481--1512.

\bibitem[Imbens and Rubin, 2015]{imbens2015causal}
Imbens, G.~W. and Rubin, D.~B. (2015).
\newblock {\em Causal inference in statistics, social, and biomedical
  sciences}.
\newblock Cambridge university press.

\bibitem[Kasy, 2014]{kasy2014instrumental}
Kasy, M. (2014).
\newblock Instrumental variables with unrestricted heterogeneity and continuous
  treatment.
\newblock {\em The Review of Economic Studies}, 81(4):1614--1636.

\bibitem[Kolsrud et~al., 2018]{kolsrud2018optimal}
Kolsrud, J., Landais, C., Nilsson, P., and Spinnewijn, J. (2018).
\newblock The optimal timing of unemployment benefits: Theory and evidence from
  sweden.
\newblock {\em American Economic Review}, 108(4-5):985--1033.

\bibitem[Kosorok, 2008]{kosorok2008introduction}
Kosorok, M.~R. (2008).
\newblock {\em Introduction to empirical processes and semiparametric
  inference}.
\newblock Springer.

\bibitem[Landais, 2015]{landais2015assessing}
Landais, C. (2015).
\newblock Assessing the welfare effects of unemployment benefits using the
  regression kink design.
\newblock {\em American Economic Journal: Economic Policy}, 7(4):243--278.

\bibitem[Landais and Spinnewijn, 2021]{landais2021value}
Landais, C. and Spinnewijn, J. (2021).
\newblock The value of unemployment insurance.
\newblock {\em The Review of Economic Studies}, 88(6):3041--3085.

\bibitem[Qu and Yoon, 2015]{qu2015nonparametric}
Qu, Z. and Yoon, J. (2015).
\newblock Nonparametric estimation and inference on conditional quantile
  processes.
\newblock {\em Journal of Econometrics}, 185(1):1--19.

\bibitem[Qu and Yoon, 2019]{qu2019uniform}
Qu, Z. and Yoon, J. (2019).
\newblock Uniform inference on quantile effects under sharp regression
  discontinuity designs.
\newblock {\em Journal of Business \& Economic Statistics}, 37(4):625--647.

\bibitem[Rubin, 2005]{rubin2005causal}
Rubin, D.~B. (2005).
\newblock Causal inference using potential outcomes: Design, modeling,
  decisions.
\newblock {\em Journal of the American statistical Association},
  100(469):322--331.

\bibitem[Sasaki, 2015]{sasaki2015quantile}
Sasaki, Y. (2015).
\newblock What do quantile regressions identify for general structural
  functions?
\newblock {\em Econometric Theory}, 31(5):1102--1116.

\bibitem[Van~der Vaart, 2000]{van2000asymptotic}
Van~der Vaart, A.~W. (2000).
\newblock {\em Asymptotic statistics}, volume~3.
\newblock Cambridge university press.

\end{thebibliography}

\end{document}